\documentclass[a4paper,11pt]{article}
\usepackage{amsmath}
\usepackage{braket}
\usepackage{amssymb}
\usepackage{slashed}
\pdfoutput=1 

\usepackage{jheppub} 

\usepackage[T1]{fontenc} 

\newcommand{\cM}{{\cal M}}

\usepackage{ulem}

\allowdisplaybreaks

\title{\boldmath One-loop Lipatov vertex in QCD with higher $\epsilon$-accuracy}

\author[a,b]{Victor S. Fadin,}
\author[c,d,e,1]{Michael Fucilla,\note{Corresponding author.}}
\author[c,d]{Alessandro Papa}

\affiliation[a]{ Budker Institute of Nuclear Physics of SB RAS, 630090 Novosibirsk, Russia}
\affiliation[b]{Novosibirsk State University, 630090 Novosibirsk, Russia}
\affiliation[c]{Dipartimento di Fisica, Universit\`a della Calabria, I-87036 Arcavacata di Rende, Cosenza, Italy}
\affiliation[d]{Istituto Nazionale di Fisica Nucleare, Gruppo collegato di Cosenza, I-87036 Arcavacata di Rende, Cosenza, Italy}
\affiliation[e]{Université Paris-Saclay, CNRS, IJCLab, 91405 Orsay, France}

\emailAdd{fadin@inp.nsk.su}
\emailAdd{michael.fucilla@unical.it}
\emailAdd{alessandro.papa@fis.unical.it}

\abstract{The effective Reggeon-Reggeon-gluon vertex, known as Lipatov vertex, is the key ingredient that allows to develop the BFKL approach in QCD. Within the next-to-leading logarithmic approximation, it is sufficient to know its one-loop corrections, in dimensional regularization ($D=4+2\epsilon$), up to the constant term in the $\epsilon$-expansion. In the next-to-next-to-leading approximation, however, the one-loop Lipatov vertex is needed up to the order $\epsilon^2$. In this paper we present the expression for this vertex in dimensional regularization up to the required accuracy.}


\begin{document} 
\maketitle
\flushbottom

\clearpage

\section{Introduction}
The famous BFKL  equation~\cite{Fadin:1975cb,Balitsky:1978ic}, which determines the behaviour  of QCD amplitudes  at high energies, is based on the gluon Reggeization.  In the BFKL approach, discontinuities of elastic amplitudes are given by the convolution of the Green's function of two interacting Reggeized gluons with the impact factors of colliding particles, the latter describing the interaction of these particles with Reggeized gluons. The BFKL equation describes the evolution with energy of the Green's function.  The kernel of the  equation consists of two parts. The first one, which is called {\em virtual}, is given by the sum of the trajectories of Reggeons. The second part, called {\em real}, is represented by convolutions of vertices that describe the emission of particles in Reggeon-Reggeon interactions. In the leading logarithmic approximation (LLA)  only a gluon can be emitted in Reggeon-Reggeon collisions. The real part of BFKL kernel at leading order (LO), entering the LLA BFKL equation, takes contribution just from the  Reggeon-Reggeon-gluon (RRG) vertex, $\gamma_{R_1R_2}^G$, in the Born approximation.  It was calculated in \cite{Lipatov:1976zz} and is usually called {\em Lipatov vertex}.  In  the next-to-leading logarithmic approximation (NLLA), one-loop corrections to this vertex are necessary.  There are gluon and quark corrections, coming from gluon and quark loops, correspondingly. 
The corrections contain infrared divergences, which are regularized taking the space-time dimension  $D=4+2\epsilon$.  Of course, gluon corrections are the most complicated ones. They were calculated firstly in~\cite{Fadin:1993rc}, where only terms finite at  $\epsilon\rightarrow 0$ were kept.  But later it was realised that,  because of the singular behaviour of the  kernel at small values of the transverse momentum $\vec p$ of the produced gluon,  the RRG vertex at small  $\vec p$ must be known with accuracy of order $\epsilon$ for a soft gluon, {\it i.e.} in the region $\epsilon \ln (1/\vec p^{\; 2})\sim 1$.  Thereafter, the vertex in the region of such small $\vec p$ values was calculated~\cite{Fadin:1996yv} at arbitrary $D$. These results were confirmed in \cite{DelDuca:1998cx}, where the vertex was obtained with accuracy of order $\epsilon^0$ in general kinematics and $\epsilon^1$ in the small-$\vec p$ region, using the results of~\cite{Bern:1993mq}.  The quark part of the one-loop correction to the vertex was calculated~\cite{Fadin:1994fj} at arbitrary $D$ from the beginning. In supersymmetric generalizations, the BFKL kernel contains also the contribution of scalar particles. This part was found in~\cite{Gerasimov:2010zzb} also at arbitrary $D$. 
 
At present, the BFKL kernel in the next-to-the-leading order (NLO) is known both for forward scattering, {\it i.e.} $t = 0$ and vacuum quantum numbers in the $t$-channel~\cite{Fadin:1998py,Ciafaloni:1998gs}, and for $t\neq 0$ and any possible two-gluon color exchange in the $t$-channel~\cite{Fadin:1998jv,Fadin:2000kx,Fadin:2000hu,Fadin:2005zj}. The kernel for the forward scattering is widely used in phenomenology (see, for example,~\cite{Ducloue:2013bva,Celiberto:2020tmb,Celiberto:2020wpk} and references therein). 

The problem of developing the BFKL approach in the next-to-NLLA (NNLLA) and, in particular, the calculation of the next-to-NLO (NNLO) corrections to the BFKL kernel, has been standing for a long time. The one-loop central emission vertex for two gluons which are not strongly ordered in rapidity in $\mathcal{N}=4$ SYM and the pure gauge, or $n_f=0$, part of the QCD three-loop Regge trajectory have been computed, respectively, in~\cite{Byrne:2022wzk} and~\cite{DelDuca:2021vjq}. Recently, the full QCD three-loop Regge trajectory has been computed in~\cite{Caola:2021izf} and~\cite{Falcioni:2021dgr}. It is also important to mention that the BFKL equation, in the LLA and in the NLLA, is derived using the pole Regge form of QCD amplitudes with gluon quantum numbers in cross-channels and negative signature. However, this form is violated in the NNLLA. This was first shown in~\cite{DelDuca:2001gu} when considering the non-logarithmic terms in two-loop amplitudes for elastic scattering. A detailed consideration of the terms responsible for the breaking of the pole-Regge form in two- and three-loop amplitudes was performed in~\cite{DelDuca:2013ara,DelDuca:2013dsa,DelDuca:2014cya}. In~\cite{Fadin:2016wso,Fadin:2017nka,Caron-Huot:2017fxr} it was shown that the observed violation of the pole-Regge form can be explained by the contributions of the three-Reggeon cuts. A procedure for disentangling the Regge cut and Regge pole in QCD in all orders of perturbation theory has been suggested in~\cite{Falcioni:2021dgr}.

The NNLLA formulation of BFKL requires not only two and three-loop calculations, but also  higher $\epsilon$-accuracy of the one-loop results. Thus, since the NLO RRG vertex has a singularity $1/\epsilon^2$, it must be known in general up to terms of order $\epsilon^2$ to ensure the accuracy  $\epsilon^0$ in the part of the kernel containing the product of two RRG vertices. Of course,  in the region of small transverse momentum $\vec p$ of the produced gluon, the accuracy must be higher ($\epsilon^3$). Fortunately, in this region the vertex was obtained in~\cite{Fadin:1996yv} exactly in $\epsilon$. Therefore, an urgent task is to calculate the one-loop RRG vertex for any $\vec p$ with accuracy $\epsilon^2$. In the planar maximally supersymmetric $ \mathcal{N} = 4$ Yang-Mills theory, the Lipatov vertex, within accuracy $\epsilon^2$, has been computed in~\cite{DelDuca:2009ae}. 

Another context requiring knowledge  of the vertex with high accuracy is related with the impact factors for the  Reggeon-gluon transition, which are the natural generalization of the  impact factors for the particle-particle transitions, entering the  discontinuities of elastic amplitudes. Similarly, discontinuities  of  amplitudes of multiple gluon production  in the multi-Regge kinematics (MRK) contain Reggeon-gluon impact factors, which describe transitions of Reggeons (Reggeized gluons) into particles (ordinary gluons), due to interaction with Reggeized gluons.   These impact factors appeared  firstly~\cite{Bartels:2003jq} in the derivation of the bootstrap conditions for inelastic amplitudes. They contain the Lipatov vertex as an integral part (see \cite{Ioffe:2010zz} and references therein).  
 
We remind also that the BFKL equation is derived using $s$-channel unitarity relations for the elastic amplitudes, with  account of contributions of multiple production amplitudes in the MRK.   The  discontinuities of these amplitudes  are not required   in the LLA and NLLA, because they are  suppressed by one power of some large logarithm,  in comparison with the real parts of the amplitudes.  But  their   account in the NNLLA  is  obligatory.  Their calculation requires the knowledge of the Lipatov vertex with high accuracy. 
 
The  discontinuities of multiple gluon production  amplitudes in the MRK  are interesting also from another point of view. They can be used~\cite{Fadin:2014yxa} for a simple demonstration of  violation of the  ABDK-BDS (Anastasiou-Bern-Dixon-Kosower ---  Bern-Dixon-Smirnov) ansatz~\cite{Anastasiou:2003kj,Bern:2005iz} for amplitudes  with maximal helicity violation (MHV) in Yang-Mills theories with maximal supersymmetry ($\mathcal{N}=4$ SYM) in the planar limit and for the calculations of the remainder functions to this ansatz. There are two hypothesis about the remainder functions: the hypothesis of dual conformal invariance~\cite{Bern:2006ew,Nguyen:2007ya}, which asserts that the MHV amplitudes are given by products of the BDS amplitudes and that the remainder functions depend only on anharmonic ratios of kinematic invariants, and the hypothesis of scattering amplitude/Wilson loop correspondence~\cite{Drummond:2007aua,Drummond:2007cf,
Brandhuber:2007yx,Drummond:2008aq}, according to which the remainder functions are given by the expectation values of Wilson loops.  Both these hypotheses are not proved. They can be tested by comparison of the BFKL discontinuities with the discontinuities calculated with the use of these hypotheses~\cite{Lipatov:2010qg,Fadin:2011we}.
 
One more problem  requiring knowledge  of the vertex with high accuracy is the proof of gluon Reggeization, which appeared originally as an hypothesis, on the basis of direct calculations firstly of elastic amplitudes in two loops~\cite{Lipatov:1976zz}, then of elastic amplitudes in  three loops and of particle-production amplitudes in one loop~\cite{Kuraev:1976ge}.   The hypothesis is extremely powerful since an infinite number of amplitudes in all orders of perturbation theory is expressed in terms of the gluon Regge trajectory and several Reggeon vertices.  Evidently, its proof is extremely desirable. It was performed both in the LLA~\cite{Balitsky:1979ap} and in the NLLA (see~\cite{Fadin:2015ym} and references therein) using bootstrap relations following from the requirement of compatibility of the pole Regge form with the $s$-channel unitarity. It turns out, that an infinite number of these relations is fulfilled if several bootstrap conditions are fulfilled~\cite{Fadin:2006pr}. The fulfillment of these conditions has been shown in a number of papers (see~\cite{Fadin:2015ym} and references therein). However, not all conditions were checked at arbitrary $D$. Some of them were checked only in the limit $\epsilon \rightarrow 0$. One of the reasons for that was the inadequate accuracy of the Lipatov vertex. This shortcoming was eliminated in~\cite{Fadin:2000yp}, where the one-loop gluon correction to the vertex was obtained at arbitrary $D$, albeit in a form containing several integrals. But this form  turns out to be inconvenient for some applications. Therefore, it is desirable to have other representations of the Lipatov vertex with high accuracy in $\epsilon$. 
 
 In this paper we present the result of the calculations of NLO RRG vertex in QCD  with accuracy up to terms of the order $\epsilon^2$. The paper contains four Sections and four Appendices. In Section~2 we introduce notations, review  the results obtained in~\cite{Fadin:2000yp} and give the representation of the vertex obtained there. In the Section~3 the calculations of basic integrals entering this representation  are described. Section~4 contains our conclusions and outlook. Appendices contain usual definitions, useful integrals, relations and limits. 

\label{sec:Intro}

\section{Review of the Lipatov vertex at one-loop}
\label{sec:Rew}
\subsection{The gluon production amplitude}
\label{subsec:gluonproductionamplitude}
The RRG vertex can be obtained from a generic $A + B \rightarrow A' + g + B'$ amplitude taken in MRK and with the gluon emitted in the central kinematical region. In this case we follow the construction in~\cite{Fadin:2000yp} and consider gluon-gluon collision. We will use the denotations $p_A$ and $p_B$ ($p_{A'}$ and $p_{B'}$) for the momenta of the incoming (outgoing) gluons, $p$ and $e(p)$ for momentum and polarization vector of the produced gluon; $q_1= p_A - p_{A'}$ and $q_2= p_{B'} - p_B$ are the momentum transfers, so that $p=q_1-q_2$. The denotations are the same as in~\cite{Fadin:2000yp}, except that we use $p$ for the ``central" outgoing gluon, instead of $k$. The MRK kinematics is defined by the relations
\begin{equation}
s \gg s_1 , s_2 \gg |t_1| \sim |t_2|\;,
\label{2}
\end{equation}
where
\begin{equation}
s=(p_A+p_B)^2\;, \;\;\;\;\; s_1=(p_{A'}+p)^2\;, \;\;\;\;\; 
s_2=(p_{B'}+p)^2\;, 
\;\;\;\;\; t_{1,2}=q_{1,2}^2\;.
\label{MRKMin}
\end{equation}
In terms of the parameters of the Sudakov decomposition
\begin{equation}
p = \beta_p \, p_A + \alpha_p \, p_B + p_\perp \;, \;\;\;\;\;\;\;\;\;\;
q_i = \beta_i \, p_A + \alpha_i \, p_B + {q_i}_{\perp} \;, 
\label{4}
\end{equation}
the relations~(\ref{2}) give
\[
1 \gg \beta_p \approx \beta_1 \gg -\alpha_1 \simeq \frac{\vec q_1^{\:2}}{s} \;,
\;\;\;\;\;\;\;\;\;\;
1 \gg \alpha_p \approx -\alpha_2 \gg \beta_2 \simeq\frac{\vec q_2^{\:2}}{s} \;,
\]
\begin{equation}
s_1 \approx s \, \alpha_p \;, \;\;\;\;\; s_2 \approx s \, \beta_p \;, \;\;\;\;\;
\vec{p}^{\; 2} = - p_\perp^2 \approx \frac{s_1 s_2}{s}\;.
\label{5}
\end{equation} 
Here and below, the vector sign is used for the components of the momenta transverse to the plane of the momenta of the initial particles $p_A$ and $p_B$. \\

To extract the RRG vertex, we can restrict ourselves to amplitudes with conservation of the helicities of the scattered gluons. The form of these amplitudes is well known:
\begin{equation}
A_{2\rightarrow 3} = 2s \, g^3\, T_{A'A}^{c_1} \, \frac{1}{t_1} \, T_{c_2c_1}^{d} 
\, \frac{1}{t_2}\,T_{B'B}^{c_2} \, e^*_\mu(p) \, {\cal A}^\mu(q_2,q_1)\;,
\label{6}
\end{equation} 
where $T_{bc}^a$ are matrix elements of the colour group generator 
in the adjoint representation and the amplitude
${\cal A}^\mu$ in the Born approximation is equal to $C^\mu(q_2,q_1)$: 
\begin{equation}
{\cal A}_{\rm Born}^\mu = C^\mu(q_2,q_1) = -{q_1}_{\perp}^\mu -{q_2}_{\perp}^\mu
+\frac{p_A^\mu}{s_1}\,(\vec p^{\; 2} - 2\vec q_1^{\:2})
-\frac{p_B^\mu}{s_2}\,(\vec p^{\; 2} - 2\vec q_2^{\:2}) \; .
\label{7}
\end{equation}
It was shown in~\cite{Fadin:1993rc} that at one-loop order the amplitude can be presented
 in its gauge invariant form,
\begin{equation}
{\cal A}^\mu = C^\mu(q_2,q_1) (1+\overline g^2 r_C) + {\cal P}^\mu \overline g^2
\, 2 t_1 t_2 \, r_{\cal P}\;,
\label{26}
\end{equation}
where terms proportional to $p^\mu$ were obviously omitted and 
\begin{equation*}
    \bar{g} = \frac{N g^2 \Gamma (1-\epsilon)}{(4 \pi)^{2+\epsilon}} \; ,
\end{equation*}
\begin{equation*}
    \mathcal{P}^{\mu} = \frac{p_A^{\mu}}{s_1} - \frac{p_B^{\mu}}{s_2} \; ,
\end{equation*}
\[
r_C = \left\{t_1 t_2 \left(r_{_+}+\frac{{\cal F}_5}{2}\right)+t_2{\cal I}_{4B} 
+ \frac{\Gamma^2(\epsilon)}{\Gamma(2\epsilon)} (\vec q_1^{\:2})^\epsilon
\left[ -\frac{1}{2}\ln\left(\frac{s_1 (-s_1)}{t_1^2}\right)
+2 \psi(\epsilon) - \psi(2\epsilon)
\right.\right.
\] 
\[
- \psi(1-\epsilon)+\frac{1}{2\epsilon(1+2\epsilon)(3+2\epsilon)}\left(-3(1+\epsilon)
-\frac{\epsilon^2}{1+\epsilon} + \frac{t_1(3+14\epsilon+8\epsilon^2)
-t_2(3+3\epsilon+\epsilon^2)}{t_1-t_2} \right. 
\]
\[
\left.\left.
+ \frac{\vec p^{\; 2} \, t_1 \, \epsilon}{(t_1-t_2)^3}
\biggl((2+\epsilon)t_2-\epsilon\,t_1\biggr)\right]\right\}
+ \biggl\{A \longleftrightarrow B \biggr\}\;,
\]
\[
r_{\cal P} = \left\{(\vec q_1 \cdot \vec q_2) r_{_+} - \frac{t_1+t_2}{4}{\cal F}_5
-\frac{{\cal I}_{4B}}{2} 
+ \frac{1}{2}\frac{\Gamma^2(\epsilon)}{\Gamma(2\epsilon)} 
(\vec q_1^{\:2})^{\epsilon-1}
\left( -\frac{1}{2}\ln\left(\frac{s_1(-s_1)s_2(-s_2)}{s(-s)t_1^2}\right)
+ \psi(1)
\right.\right.
\]
\[
+\psi(\epsilon) - \psi(1-\epsilon) - \psi(2\epsilon)
+\frac{t_1}{t_2\,(1+2\epsilon)(3+2\epsilon)}
\left[ \frac{t_2}{t_1-t_2} (11+7\epsilon) \right.
\]
\begin{equation}
\left.\left.\left.
+ \frac{\vec p^{\; 2}}{(t_1-t_2)^3}\biggl(t_2(t_1+t_2)-\epsilon\,t_1(t_1-t_2)\biggr)
+ \frac{(\vec p^{\; 2})^2}{(t_1-t_2)^3}\biggl((2+\epsilon)t_2-\epsilon\,t_1\biggr)
\right]\right)\right\} + \biggl\{ A \longleftrightarrow B \biggr\} .
\label{27}
\end{equation}
The expression for the function $r_+$ appearing in the definition of ${\cal A}^\mu$ is
\begin{equation}
r_{_+} = \frac{f_{_-} (\vec q_1^{\:2}-\vec q_2^{\:2})-f_{_+} \vec p^{\; 2}}
{8 (\vec q_1^{\:2} \vec q_2^{\:2} - (\vec q_1 \vec q_2)^2)} \;,
\end{equation}
with
\[
f_{_-} \equiv \left[-t_1 {\cal F}_5 - {\cal I}_{4B}
- \frac{\Gamma^2(\epsilon)}{\Gamma(2\epsilon)}(\vec q_2^{\: 2})^{\epsilon-1}
\left(\frac{1}{2} \ln\left(\frac{s_1 (-s_1) s_2(-s_2)}{s(-s) t_2^2}\right)
\right.\right.
\]
\begin{equation}
\biggl. \biggl.
+ \psi(1-\epsilon) - \psi(\epsilon) + \psi(2 \epsilon) - \psi(1) \biggr)\biggr]
- \biggl[ A \longleftrightarrow B \biggr] \;, 
\label{21}
\end{equation}
\[
 f_{_+} \equiv \left\{-t_1 {\cal F}_5 
-{\cal I}_{4B}-\frac{\Gamma^2(\epsilon)}{\Gamma(2\epsilon)}
\left[(\vec q_2^{\: 2})^{\epsilon-1}
\left(\frac{1}{2} \ln\left(\frac{s_1 (-s_1) s_2(-s_2)}{s(-s) t_2^2}\right)
\right.\right.\right.
\]
\[
\biggl. 
+ \psi(1-\epsilon) - \psi(\epsilon) + \psi(2 \epsilon) - \psi(1) \biggr)
\]
\begin{equation}
\left.\left.
-(\vec p^{\; 2})^{\epsilon-1}
\left(\frac{1}{2} \ln\left(\frac{s_1 (-s_1) s_2(-s_2)}{s(-s) (\vec p^{\; 2})^2}
\right) + \psi(1-\epsilon) - \psi(\epsilon)\right)\right]\right\}
+ \biggl\{ A \longleftrightarrow B \biggr\} \; , 
\label{22}
\end{equation}
and
\begin{equation}
    {\cal F}_5 = {\cal I}_5 - {\cal L}_3 - \frac{1}{2} 
\ln\left(\frac{s (-s) (\vec p^{\; 2})^2}{s_1 (-s_1) s_2 (-s_2)}\right) {\cal I}_3\;.
\end{equation}
The total amplitude is given in terms of five structures $ \mathcal{I}_3,  \; \mathcal{L}_3, \; \mathcal{I}_{4B}, \; \mathcal{I}_{4A}, \; \mathcal{I}_{5}$, defined as~\cite{Fadin:2000yp}
\begin{equation}
    \mathcal{I}_3 = \int \frac{d^{2+2\epsilon} k}{\pi^{1+\epsilon}\Gamma(1-\epsilon)} \frac{1}{\vec{k}^{\, 2}  (\vec{k}-\vec{q}_1)^2 (\vec{k}-\vec{q}_2)^2} \; ,
    \label{I3cal}
\end{equation}
\begin{equation}
    \mathcal{L}_3 = \int \frac{d^{2+2\epsilon} k}{\pi^{1+\epsilon} \Gamma(1-\epsilon)} \frac{1}{\vec{k}^{\, 2}  (\vec{k}-\vec{q}_1)^2 (\vec{k}-\vec{q}_2)^2} \left[\ln \left( \frac{(\vec{k}-\vec{q}_1)^2(\vec{k}-\vec{q}_2)^2}{\vec{k}^{\, 2}  (\vec{q}_1-\vec{q}_2)^{\; 2}} \right) \right] \; ,
    \label{L3cal}
\end{equation}
\begin{equation*}
\mathcal{I}_{4B} = \int_0^1 \frac{dx}{x} \int \frac{d^{2+2\epsilon} k}{\pi^{1 + \epsilon} \Gamma (1-\epsilon)} 
\end{equation*} 
\begin{equation}
   \times \left[ \frac{1-x}{\left( x \vec{k}^2 + (1-x) (\vec{k}-\vec{q}_1)^2 \right) (\vec{k}-(1-x)(\vec{q}_1-\vec{q}_2))^2} - \frac{1}{(\vec{k}-\vec{q}_{1})^2 (\vec{k}-(\vec{q}_1-\vec{q}_2))^2} \right] \; ,
\label{I4Bcal}
\end{equation}
\begin{equation*}
    \mathcal{I}_5 = \int_0^1 \frac{d x}{1-x} \int \frac{ d^{2+2\epsilon} k}{\pi^{1+\epsilon} \Gamma(1-\epsilon)} \frac{1}{\vec{k}^{\, 2} [(1-x) \vec{k}^{\, 2}  + x (\vec{k}-\vec{q}_1)^2]}
\end{equation*}
\begin{equation}
   \times \left[ \frac{x^2}{(\vec{k}-x(\vec{q}_1-\vec{q}_2))^2} - \frac{1}{(\vec{k}-\vec{q}_1+\vec{q}_2)^2} \right] 
    \label{I5cal}
\end{equation}
and $\mathcal{I}_{4A}= \mathcal{I}_{4B} (\vec{q}_1 \leftrightarrow -\vec{q_2})$.

\subsection{The Lipatov vertex}
\label{subsec: Lipatov vertex}
Imposing general requirements of
analyticity, unitarity and crossing symmetry, the production amplitude must take the Regge form~\cite{Fadin:1993rc,Bartels:1980}
\[
8\, g^3 \, {\cal A}^\mu = \Gamma(t_1;\vec p^{\; 2}) \left\{
\left[\left(\frac{s_1}{\vec p^{\; 2}}\right)^{\omega_1-\omega_2}
+\left(\frac{-s_1}{\vec p^{\; 2}}\right)^{\omega_1-\omega_2}\right]
\left[\left(\frac{s}{\vec p^{\; 2}}\right)^{\omega_2}
+\left(\frac{-s}{\vec p^{\; 2}}\right)^{\omega_2}\right]\, R^\mu \right.
\]
\begin{equation}
\left.
+ \left[\left(\frac{s_2}{\vec p^{\; 2}}\right)^{\omega_2-\omega_1}
+\left(\frac{-s_2}{\vec p^{\; 2}}\right)^{\omega_2-\omega_1}\right]
\left[\left(\frac{s}{\vec p^{\; 2}}\right)^{\omega_1}
+\left(\frac{-s}{\vec p^{\; 2}}\right)^{\omega_1}\right]\, L^\mu 
\right\}\Gamma(t_2;\vec p^{\; 2}) \;,
\label{28}
\end{equation}
where $\omega_i = \omega(t_i)$ (we have chosen $\vec p^{\; 2}$ as the scale of energy), $\Gamma(t_i;\vec p^{\; 2})$ are the helicity conserving gluon-gluon-Reggeon vertices, $R^\mu$ and $L^\mu$ are the right and left 
RRG vertices, depending on $\vec q_1$ and $\vec q_2$. They are real 
in all physical channels, as well as $\Gamma(t_i; \vec p^{\; 2})$. 

In the one-loop approximation, we have 
\begin{equation}
\omega(t) = \omega^{(1)}(t) = - \overline g^2 \frac{\Gamma^2(\epsilon)}
{\Gamma(2\epsilon)} (\vec q^{\:2})^\epsilon\;,
\label{29}
\end{equation}
\begin{equation}
\Gamma(t;\vec p^{\; 2}) =  g \biggl(1+\Gamma^{(1)}(t;\vec p^{\; 2}) \biggr)\;,
\label{30}
\end{equation}
where
\[
\Gamma^{(1)}(t;\vec p^{\; 2}) = \overline g^2 \frac{\Gamma^2(\epsilon)}
{\Gamma(2\epsilon)} (\vec q^{\:2})^\epsilon \left[\psi(\epsilon)
- \frac{1}{2} \psi(1) - \frac{1}{2} \psi(1-\epsilon) \right.
\]
\begin{equation}
\left.
+\frac{9 (1+\epsilon)^2+2}{4(1+\epsilon) (1+2\epsilon) (3+2\epsilon)}
-\frac{1}{2} \ln\left(\frac{\vec p^{\; 2}}{\vec q^{\:2}}\right)\right] \;.
\label{31}
\end{equation}
In the same approximation, we obtain from~(\ref{28})
\[
2 g \left\{{\cal A}^\mu - C^\mu(q_2,q_1)\left[\Gamma^{(1)}(t_1;\vec p^{\; 2})
+\Gamma^{(1)}(t_2;\vec p^{\; 2})
+\frac{\omega_1}{2}\ln\left(\frac{s_1(-s_1)}{(\vec p^{\; 2})^2}\right)
+\frac{\omega_2}{2}\ln\left(\frac{s_2(-s_2)}{(\vec p^{\; 2})^2}\right) \right.\right.
\]
\begin{equation}
\left.\left.
+\frac{\omega_1+\omega_2}{4}\ln\left(\frac{s(-s)(\vec p^{\; 2})^2}{s_1(-s_1)s_2(-s_2)}
\right)\right]\right\}
= R^\mu + L^\mu + (R^\mu - L^\mu) \frac{\omega_1-\omega_2}{4} 
\ln\left(\frac{s_1(-s_1)s_2(-s_2)}{s(-s)(\vec p^{\; 2})^2}\right) .
\label{32}
\end{equation}
Since in all physical channels
\begin{equation}
    \ln\left(\frac{s_1(-s_1)s_2(-s_2)}{s(-s)(\vec p^{\; 2})^2}\right) = i \pi \;,
\end{equation}
the combinations $R^{\mu} + L^{\mu}$ and $R^{\mu} - L^{\mu}$ are respectively related to the real and imaginary parts of the production amplitude. Using this, we find that
\[
R^\mu + L^\mu = 2g\left\{C^\mu(q_2,q_1)+
\overline g ^2\left(\frac{(\vec q_1 \cdot \vec q_2)C^\mu(q_2,q_1)+
2\vec q_1^{\:2}\vec q_2^{\:2}{\cal P}^\mu}{2(\vec q_1^{\:2}\vec q_2^{\:2}
-(\vec q_1 \cdot \vec q_2)^2)}\left[
\frac{\vec q_1^{\:2}\vec q_2^{\:2}\vec{p}^{\; 2}}{2}
({\cal I}_5-{\cal L}_3)
\right.\right.\right.
\]
\[
\left.
-\vec q_2^{\:2}(\vec q_1 \cdot \vec{p})\,{\cal I}_{4B}
+\frac{\Gamma^2(\epsilon)}{\Gamma(2\epsilon)}
\left(\frac{(\vec{p}^{\; 2})^{\epsilon}}{2}(\vec q_1 \cdot \vec q_2)\left(
\psi(\epsilon)-\psi(1-\epsilon)\right)+(\vec q_1^{\:2})^{\epsilon}
(\vec q_2 \cdot \vec{p})\left(\ln \left(\frac{\vec p^{\; 2}}
{\vec q_1^{\:2}}\right)-\psi(1)\right.\right.\right.
\]
\[
\left.\left.\left. -\psi(\epsilon) +\psi(1-\epsilon)
+\psi(2\epsilon)\phantom{\frac{1}{1}}\!\!\!
\right)\right)\right]+C^\mu(q_2,q_1)\left[-\frac{\vec q_2^{\:2}}{2}{\cal I}_{4B}+
\frac{\Gamma^2(\epsilon)}{2\Gamma(2\epsilon)}\left(
\frac{(\vec{p}^{\; 2})^\epsilon}{2}(\psi(\epsilon)-\psi(1-\epsilon))
\right.\right.
\]
\[
+(\vec q_1^{\:2})^{\epsilon}\left[\psi(\epsilon)
-\psi(2\epsilon)+\frac{1}{2(1+2\epsilon)(3+2\epsilon)}\left(\frac
{\vec q_1^{\:2}+\vec q_2^{\:2}}{\vec q_1^{\:2}-\vec q_2^{\:2}}
(11+7\epsilon)+2\epsilon\frac
{\vec p^{\; 2}\vec q_1^{\:2}}{(\vec q_1^{\:2}-\vec q_2^{\:2})^2} \right.\right.
\]
\[
\left.\left.\left.\left.
-4\frac{\vec p^{\; 2}\vec q_1^{\:2}\vec q_2^{\:2}}
{(\vec q_1^{\:2}-\vec q_2^{\:2})^3}\right)\right]\right)\right]
+{\cal P}^\mu\frac{\Gamma^2(\epsilon)(\vec q_1^{\:2})^{\epsilon}}
{\Gamma(2\epsilon)(1+2\epsilon)(3+2\epsilon)}
\left[\frac{\vec q_1^{\:2}\vec q_2^{\:2}}{\vec q_1^{\:2}-\vec q_2^{\:2}}
(11+7\epsilon)+\epsilon \vec p^{\; 2}\vec q_1^{\:2}\frac
{\vec q_1^{\:2}-\vec p^{\; 2}}{(\vec q_1^{\:2}-\vec q_2^{\:2})^2}
\right. 
\]
\begin{equation}
\left.\left.\left.
-\vec p^{\; 2}\vec q_1^{\:2}\vec q_2^{\:2}
\frac{\vec q_1^{\:2}+\vec q_2^{\:2}-2\vec p^{\; 2}}
{(\vec q_1^{\:2}-\vec q_2^{\:2})^3}\right]\right)
-\overline g ^2\left(A\longleftrightarrow B\right)\right\}  
\label{Rmu+Lmu}
\end{equation}
and
\[
R^\mu - L^\mu =  \frac{2g \overline g^2 }{\omega_1-\omega_2}
\left\{-C^\mu(q_2,q_1)\frac{\Gamma^2(\epsilon)}
{\Gamma(2\epsilon)}(\vec p^{\; 2})^{\epsilon}
+\frac{(\vec q_1 \cdot \vec q_2)C^\mu(q_2,q_1)+2\vec q_1^{\:2}
\vec q_2^{\:2}{\cal P}^{\mu}}{\vec q_1^{\:2} \vec q_2^{\:2}
-(\vec q_1 \cdot \vec q_2)^2}
\right.
\]
\begin{equation}
\left.\times \left[\vec q_1^{\:2}\vec q_2^{\:2}\vec p^{\; 2}
{\cal I}_3+\frac{\Gamma^2(\epsilon)}{\Gamma(2\epsilon)}
\left((\vec q_1^{\:2})^{\epsilon}(\vec q_2 \cdot \vec p) 
-(\vec q_2^{\:2})^{\epsilon}(\vec q_1 \cdot \vec p) 
-(\vec p^{\; 2})^{\epsilon}(\vec q_1 \cdot \vec q_2)\right)\right]\right\}\;.
\label{Rmu-Lmu}
\end{equation} 
It is clear that, in order to know the Lipatov vertex at a certain order in the $\epsilon$-expansion, we must calculate the integrals $\mathcal{I}_3$, $\mathcal{I}_{4A}$, $\mathcal{I}_{4B}$, and $\mathcal{I}_5-\mathcal{L}_3$ with the same accuracy.  
\\

In the region of small momenta of the central emitted gluon (the {\em soft region}, from now on) the vertex must be known to all orders in $\epsilon$. The soft limit of the integrals entering the vertex is computed in Appendix~\ref{Appendix D}. By using Eqs.~(\ref{I3calSoft}),~(\ref{I4BcalSoft}) and (\ref{I5cal-L3calSoft}), we easily find that, in the soft limit,
\begin{equation}
R^\mu + L^\mu = 2g C^\mu(q_2,q_1)\left(1+\overline g^2
\frac{\Gamma^2(\epsilon)}{2\Gamma(2\epsilon)}(\vec p^{\; 2})^{\epsilon}
\left[\psi(\epsilon)-\psi(1-\epsilon)\right]\right), 
\label{48b}   
\end{equation}
\begin{equation}
R^\mu - L^\mu = - \frac{2g \overline g^2 }{\omega_1-\omega_2}
C^\mu(q_2,q_1)\frac{\Gamma^2(\epsilon)}
{\Gamma(2\epsilon)}(\vec p^{\; 2})^{\epsilon} \; . 
\label{36c}   
\end{equation}
The last two equation confirm the result in~\cite{Fadin:1996yv,DelDuca:1998cx}.

\section{Fundamental integrals for the one-loop RRG vertex}
\label{sec:Fund}

In this Section we present the result at $\epsilon^2$-accuracy of the integrals in the transverse momentum space entering the one-loop RRG vertex, expanding expressions
already available in the literature and, whenever possible, cross-checking them with an alternative calculation. 

\subsection{$\mathcal{I}_3$: Bern-Dixon-Kosower (BDK) method}
\label{subsec:I31}

We start from the integral
\[
\mathcal{I}_3' \equiv \pi^{1+\epsilon} \Gamma(1-\epsilon)\ \mathcal{I}_3 
\]
\begin{equation}
= \int d^{2+2\epsilon}k \frac{1}{\vec{k}^{2} (\vec{k}-\vec{q}_{1})^{2} (\vec{k}-\vec{q}_{2})^{2}} = \int d^{2+2\epsilon}k_{E} \frac{1}{k_{E}^{2} (k_{E}-q_{1E})^{2} (k_{E}-q_{2E})^{2}} \; ,
\label{I3prime_def}
\end{equation}
where $k_{E}, q_{1E}, q_{2E}$ are Euclidean vectors in dimension $2+2\epsilon$ and $q_{1E}^{\;2}, q_{2E}^{\;2}, (q_{1E}-q_{2E})^2 \neq 0$. We can relate this Euclidean integral to a corresponding Minkowskian integral by the Wick rotation:
\begin{equation}
  \int d^{2+2 \epsilon}k_{E} \frac{1}{k_{E}^{2} (k_{E}-q_{1E})^{2} (k_{E}-q_{2E})^{2}}
  = i \int d^{2+2 \epsilon}k \frac{1}{k^{2} (k-q_1)^{2} (k-q_2)^{2}}\;,
\end{equation}
where $k = (i k_{E}^0, k_{E}^1)$, $q_1 = (i q_{1E}^0, q_{1E}^1)$, $q_2
= (i q_{2E}^0, q_{2E}^1)$. We note that
\begin{equation}
\hspace{-0.1 cm}  q_1^{\;2} = - q_{1E}^{\;2}=-\vec q_1^{\;2}, \hspace{0.4 cm} q_2^{\;2} = - q_{2E}^{\;2}=-\vec q_2^{\;2}\;,
  \hspace{0.4 cm}
  (q_1-q_2)^2 = -(q_{1E}-q_{2E})^2 =- (\vec q_1- \vec q_2)^{\;2} = - \vec{p}^{\; 2} .
\end{equation}
Hence, we need a triangular integral with three massive external legs. This
result to all order in $\epsilon$-expansion can be found in~\cite{Bern_1994},
taking care to apply the replacement ($ \epsilon \rightarrow -\epsilon + 1$),
because they calculated the integral in $d^{4 - 2 \epsilon}k$, instead we need it
in $d^{2 + 2 \epsilon}k$. One finds that
\begin{equation}
\mathcal{I}_3' = \pi^{1+\epsilon} \alpha_1 \alpha_2 \alpha_3 \left( -\frac{1}{2} \frac{\Gamma(2-\epsilon) \Gamma^2(\epsilon)}{\Gamma(2 \epsilon -1)} \right) \frac{\hat{\Delta}_3^{1/2-\epsilon}}{(1-\epsilon)^2} \left[ f \left( \delta_1 \right) + f \left( \delta_2 \right) + f \left( \delta_3 \right) + c \right] \;,
\end{equation} 
where
\begin{equation}
  \alpha_1 = \sqrt{\frac{\vec{q}_{2}^{\; 2}}{\vec{q}_{1}^{\;2} (\vec{q}_{1}-\vec{q}_{2})^2}}\;,\hspace{0.5 cm}
  \alpha_2 = \sqrt{\frac{(\vec{q}_{1}-\vec{q}_{2})^2}{\vec{q}_{1}^{\; 2} \vec{q}_{2}^{\;2}}}\;, \hspace{0.5 cm}
  \alpha_3 = \sqrt{\frac{\vec{q}_{1}^{\; 2}}{\vec{q}_{2}^{\; 2} (\vec{q}_{1}-\vec{q}_{2})^2}}\;,
\end{equation}
\begin{equation}
  \gamma_1 = -\alpha_1 + \alpha_2 + \alpha_3\;, \hspace{0.5 cm}
  \gamma_2 = \alpha_1 - \alpha_2 + \alpha_3 \;, \hspace{0.5 cm}
  \gamma_3 = \alpha_1 + \alpha_2 - \alpha_3 \;,
\end{equation}
\begin{equation}
  \delta_1 = \frac{\gamma_1}{\sqrt{\hat{\Delta}_3}}\;, \hspace{0.5 cm}
  \delta_2 = \frac{\gamma_2}{\sqrt{\hat{\Delta}_3}}\;, \hspace{0.5 cm}
  \delta_3 = \frac{\gamma_3}{\sqrt{\hat{\Delta}_3}} \;,
\end{equation}
\begin{equation}
  \hat{\Delta}_3 = - \alpha_1^2 - \alpha_2^2 - \alpha_3^2 + 2 \alpha_1 \alpha_2
  + 2 \alpha_2 \alpha_3 + 2 \alpha_1 \alpha_3  \;,
\end{equation}
\begin{equation}
  c = -2 \pi (1-\epsilon) \frac{\Gamma(2 \epsilon-1)}{\Gamma^2(\epsilon)}
  \;,
\end{equation}
\begin{equation}
\begin{split}
f \left( \delta \right) = & \frac{1}{i} \left[ \left( \frac{1+i\delta}{1-i\delta} \right)^{1-\epsilon} \; _2 F_1 \left(2 - 2 \epsilon, 1 - \epsilon, 2 - \epsilon; -\left( \frac{1+i\delta}{1-i\delta} \right) \right) \right. \\ &
   \left. - \left( \frac{1-i\delta}{1+i\delta} \right)^{1-\epsilon} \; _2 F_1 \left(2 - 2 \epsilon, 1 - \epsilon, 2 - \epsilon; -\left( \frac{1-i\delta}{1+i\delta} \right) \right) \right] \;. 
\end{split}
\end{equation}
Using the expansion (\ref{HyperExp1}) and the definitions
\begin{equation}
    z_i = - \left( \frac{1+ i \delta_i}{1-i \delta_i} \right) \; ,
\end{equation}
we find 
\begin{equation*}
    \mathcal{I}_3' = \pi^{1+\epsilon} \Gamma (1-\epsilon) \alpha_1 \alpha_2 \alpha_3 \hat{\Delta}_3^{1/2-\epsilon} \frac{ \Gamma^2(1+\epsilon)}{\Gamma(1+2 \epsilon)} \frac{ (1-2 \epsilon) }{(1-\epsilon)} \Bigg \{ \frac{\alpha_1 + \alpha_2 + \alpha_3}{\sqrt{\hat{\Delta}_3} \epsilon} + \Bigg( \pi + \sum_{i=1}^{3} \Bigg[ \frac{i \; z_i}{1-z_i} 
\end{equation*}
\begin{equation*}
    \times \left( 1 - \ln \left( -z_i \right) + \frac{1+z_i}{z_i} \ln \left( 1-z_i \right) \right) - (z_i \rightarrow z_{i}^{-1}) \Bigg] \Bigg) + \left( \pi + \sum_{i = 1}^{3} \left[ \frac{ i \; z_i}{1-z_i} \right. \right.
\end{equation*}
\begin{equation*}
    \left. \left. \times \left( 2 + \left( \frac{1+z_i}{z_i} \right) \ln \left( 1-z_i \right) \ln \left( (1-z_i)e \right) + \frac{1}{2} \ln^2 \left(-z_i \right) - \frac{1-z_i}{z_i} {\rm{Li}}_2 \left(z_i \right) \right. \right. \right.
\end{equation*}
\begin{equation*}
    \left. \left. \left. - \ln \left( -z_i \right) \left( 1 + \frac{1+z_i}{z_i} \ln \left( 1-z_i \right) \right) \right) - (z_i \rightarrow z_{i}^{-1}) \right] \right) \epsilon + \bigg( \pi (2 + \zeta (2)) 
\end{equation*}
\begin{equation*}
    \left. + \sum_{i}^{3} \left[ \frac{i \; z_i}{1-z_i} \left( 4 + \ln (1-z_i) \bigg( \frac{2(1+z_i)}{z_i} + \frac{2(1-z_i)}{z_i} \zeta (2) + \frac{\ln (1-z_i)}{3 z_i} \bigl( 3 (1+z_i) \right. \right. \right. \bigr.
\end{equation*}
\begin{equation*}
       \bigl. \left. \left. \left. + 2 (1+z_i) \ln (1-z_i) -3 (1-z_i) \ln z_i \right) \bigr) - \frac{(1-z_i)}{z_i} (1+2 \ln (1-z_i)) {\rm{Li}}_2 (z_i) - \frac{2 (1-z_i)}{z_i} \right. \right. 
\end{equation*}
\begin{equation*}
      \left. \left. \left. \times \left( {\rm{Li}}_3 (1-z_i) + \frac{{\rm{Li}}_3 (z_i)}{2} - \zeta(3) \right) - \frac{\ln^3 (-z_i)}{6} + \frac{\ln^2 (-z_i)}{2} \left( 1 + \left( \frac{1+z_i}{z_i} \right) \ln (1-z_i)  \right)  \right. \right. \right.
\end{equation*}
\begin{equation}
  \left. \left. \left. \hspace{-0.15 cm} - \ln (-z_i) \left( 2 + \left( \frac{1+z_i}{z_i} \right) \ln \left( 1-z_i \right) \ln \left( (1-z_i)e \right) - \frac{1-z_i}{z_i} {\rm{Li}}_2 \left(z_i \right) \right) \right) - (z_i \rightarrow z_{i}^{-1}) \phantom{\frac{1}{1}}\!\!
    \right]  \right) \epsilon^2 \Bigg \} \; .  
\end{equation}
Please note that $\delta_i \rightarrow - \delta_i$ is equivalent to $z_i \rightarrow z_i^{-1}$. We also observe that $\alpha_2 = \alpha_1 (\vec{q}_2 \leftrightarrow \vec{q}_1-\vec{q}_2)$ and $\alpha_3 = \alpha_1 (\vec{q}_1 \leftrightarrow \vec{q}_2)$. From this, we realize that the second and the third term of the summation can be obtained from the first by two simple substitutions. We can rewrite:
\begin{equation}
    \begin{split}
        & \mathcal{I}_3' = \pi^{1+\epsilon} \Gamma (1-\epsilon) \alpha_1 \alpha_2 \alpha_3 \hat{\Delta}_3^{1/2-\epsilon} \frac{ \Gamma^2(1+\epsilon)}{\Gamma(1+2 \epsilon)} \frac{ (1-2 \epsilon) }{(1-\epsilon)} \hat{\mathcal{S}} \left \{ \frac{\alpha_1}{\sqrt{\hat{\Delta}_3} \epsilon} + \Bigg( \hspace{-0.05 cm} \pi \hspace{-0.05 cm} + \hspace{-0.05 cm} \Bigg[ \frac{i \; z_1}{1-z_1} \left( 1 - \ln \left( -z_1 \right) \right. \right. \\ & \left. \left. + \frac{1+z_1}{z_1} \ln \left( 1-z_1 \right) \hspace{-0.05 cm} \right) \hspace{-0.05 cm} - \hspace{-0.05 cm} (z_1 \rightarrow z_{1}^{-1}) \Bigg] \hspace{-0.05 cm} \Bigg) \hspace{-0.05 cm} + \hspace{-0.05 cm} \left( \hspace{-0.05 cm} \pi \hspace{-0.1 cm} + \left[ \frac{ i \; z_1}{1-z_1} \hspace{-0.05 cm} \left( \hspace{-0.05 cm} 2 + \hspace{-0.05 cm} \left( \hspace{-0.05 cm} \frac{1+z_1}{z_1} \hspace{-0.05 cm} \right) \hspace{-0.05 cm} \ln \left( 1-z_1 \right) \ln \left( (1-z_1)e \right) \right. \right. \right. \right. \\ & \left. \left. \left. \left. + \frac{1}{2} \ln^2 \left(-z_1 \right) - \frac{1-z_1}{z_1} {\rm{Li}}_2 \left(z_1 \right) - \ln \left( -z_1 \right) \left( 1 + \frac{1+z_1}{z_1} \ln \left( 1-z_1 \right) \right) \right) - (z_1 \rightarrow z_{1}^{-1}) \right]  \right) \epsilon  \right. \\ &  + \left( \pi (2 + \zeta (2)) + \left[ \frac{i \; z_1}{1-z_1} \left( 4 + \ln (1-z_1) \left( \frac{2(1+z_1)}{z_1} + \frac{2(1-z_1)}{z_1} \zeta (2) + \frac{\ln (1-z_1)}{3 z_1}  \right. \right. \right. \right. \\ & \left. \left. \left. \times \bigg ( 3 (1+z_1) + 2 (1+z_1) \ln (1-z_1)  -3 (1-z_1) \ln z_1 \bigg ) \bigg ) - \frac{(1-z_1)}{z_1} (1+2 \ln (1-z_1)) {\rm{Li}}_2 (z_1) \right. \right. \right. \\ & \left. \left. \left. \hspace{-0.3 cm} \left. - \frac{2 (1-z_1)}{z_1} \hspace{-0.05 cm} \left( \hspace{-0.05 cm} {\rm{Li}}_3 (1-z_1) \hspace{-0.05 cm} + \hspace{-0.05 cm} \frac{{\rm{Li}}_3 (z_1)}{2} \hspace{-0.05 cm} - \hspace{-0.05 cm} \zeta(3) \hspace{-0.05 cm} \right) \hspace{-0.05 cm} - \hspace{-0.05 cm} \frac{\ln^3 (-z_1)}{6} + \frac{\ln^2 (-z_1)}{2} \left( \hspace{-0.05 cm} 1 \hspace{-0.05 cm} + \hspace{-0.05 cm} \left( \frac{1+z_1}{z_1} \right) \ln (1-z_1)  \right) \right. \right. \right. \right. \\ &  \left. - \ln (-z_1) \left( 2 + \left( \hspace{-0.05 cm} \frac{1+z_1}{z_1} \hspace{-0.05 cm} \right)  \ln \left( 1-z_1 \right) \ln \left( (1-z_1)e \right) - \frac{1-z_1}{z_1} {\rm{Li}}_2 \left(z_1 \right) \hspace{-0.1 cm} \right) \hspace{-0.05 cm} \right) \left. \hspace{-0.1 cm} - ( z_1 \rightarrow z_1^{-1} ) \bigg]  \bigg) \epsilon^2 \right \} ,
    \end{split}
    \label{I3BDKe^2}
\end{equation}
where the operator $\hat{S}$ acts on a generic function  $f(\vec{q}_{1}^{\; 2}, \vec{q}_{2}^{\; 2}, \vec{p}^{\; 2} )$ as
\begin{equation}
\hat{S} \left \{ f \left( \vec{q}_{1}^{\; 2}, \vec{q}_{2}^{\; 2}, \vec{p}^{\; 2} \right) \right \} = f \left( \vec{q}_{1}^{\; 2}, \vec{q}_{2}^{\; 2}, \vec{p}^{\; 2} \right) + f \left( \vec{q}_{1}^{\; 2}, \vec{p}^{\; 2}, \vec{q}_2^{\; 2} \right) + f \left( \vec{q}_{2}^{\; 2}, \vec{q}_{1}^{\; 2}, \vec{p}^{\; 2} \right) \; .
\end{equation}
 
\subsection{$\mathcal{I}_3$: Alternative calculation}
\label{subsec:I32}
We verify the previous result through a different calculation procedure. We introduce the Feynman parametrization to get
\begin{equation}
    \mathcal{I}_3 = (1-\epsilon) \int_0^1 dx \int_0^1 dy \frac{y^{\epsilon-1}}{(A-By)^{2-\epsilon}} \;,
\end{equation}
where $A=x \vec{q}_1^{\; 2} + (1-x) \vec{q}_2^{\; 2}$ and $B=A-x(1-x) \vec{p}^{\; 2}$. Now, it is convenient to perform the decomposition
\begin{equation*}
    \int_0^1 dx \int_0^1 dy \frac{y^{\epsilon-1}}{(A-By)^{2-\epsilon}} = \int_0^1 dy \frac{y^{\epsilon}}{(A-By)^{-\epsilon}}
\left[\frac{1/A^2}{y} +\frac{B/A^2}{A-By}+\frac{B/A}{(A-By)^2}\right]
\end{equation*}
\begin{equation*}
    = \int_0^1 dy \left[ \frac{y^{\epsilon-1}}{A^{2-\epsilon}} \left( 1 - \frac{B}{A} y \right)^{\epsilon} + \frac{B y^{\epsilon}}{A^2 (A-By)^{1-\epsilon}} + \frac{B y^{\epsilon}}{A (A-By)^{2-\epsilon}} \right]\;.
\end{equation*}
After this, we can safely expand $\left( 1 - \frac{B}{A} y \right)^{\epsilon}$ in the first term and $y^{\epsilon}$ in the second and third terms, getting
\begin{equation*}
    \mathcal{I}_3 = (1-\epsilon) \int_0^1 dx \int_0^1 dy \left \{ \frac{y^{\epsilon-1}}{A^{2-\epsilon}} + \frac{B}{A^2 (A-By)^{1-\epsilon}} + \frac{B}{A (A-By)^{2-\epsilon}} + \epsilon \left( \frac{\ln \left( 1- \frac{B}{A} y \right)}{A^2 y} \right. \right.
\end{equation*}
\begin{equation*}
    \left. \left. + \frac{B \ln y}{A^2 (A-By)} + \frac{B \ln y}{A (A-By)^2} \right) + \epsilon^2 \left[ \frac{\ln (1- \frac{B}{A} y)}{y A^2} \left( \frac{\ln (1- \frac{B}{A} y)}{2} + \ln A + \ln y \right) \right. \right.
\end{equation*}
\begin{equation*}
        \left. \left. + \frac{B \ln y}{A (A-By)}  \left( \frac{\ln y}{2} + \ln (A-By) \right) \left( \frac{1}{A} + \frac{1}{A-By} \right) \right] \right \} + {\cal O}(\epsilon^3) \;.
\end{equation*}
Integrating over $y$, we obtain
\begin{equation*}
\mathcal{I}_3 = \int_0^1 dx \left \{ \frac{(2-3 \epsilon)}{\epsilon} \frac{1}{A^{2-\epsilon}} - \frac{(1- \epsilon)}{\epsilon} \frac{(A-B)^{\epsilon}}{A^{2}} + \frac{1}{A(A-B)^{1-\epsilon}} + \epsilon \left( \frac{1}{A^2} \ln \left( \frac{A-B}{A} \right) \right. \right.
\end{equation*}
\begin{equation*}
    \left. \left. - \frac{2}{A^2} {\rm{Li}}_2 \left( \frac{B}{A} \right) \right) \right \} + 2 \epsilon^2 \int_{0}^1  \frac{dx}{A^2} \left[ 2 \; {\rm{Li}}_2 \left( \frac{B}{A} \right) + \; {\rm{S}}_{1,2} \left( \frac{B}{A} \right) \right. 
\end{equation*}
\begin{equation}
   \left. + {\rm{Li}}_3 \left( \frac{B}{A} \right) + \frac{\ln A}{2} \left( \ln \left( \frac{A-B}{A} \right) -2 \; {\rm{Li}}_2 \left( \frac{B}{A} \right)  \right) + \frac{1}{4} \ln^2 \left( \frac{A-B}{A} \right) \right] + {\cal O}(\epsilon^3)  \;.
\label{I3final}
\end{equation}

In this form, all divergences are contained in the first three terms, which we can promptly compute, to get
\begin{equation*}
     \int_0^1 dx \frac{(2-3 \epsilon)}{\epsilon A^{2-\epsilon} } = \frac{2}{ab} \left[ \frac{1}{\epsilon} + \frac{a \ln b - b \ln a}{a-b} - \frac{1}{2} + \frac{\epsilon}{2} \left( \frac{a \ln^2 b - b \ln^2 a}{a-b} -\frac{a \ln b - b \ln a}{a-b}  -1 \right) \right. 
\end{equation*}
\begin{equation}
   \left. + \frac{\epsilon^2}{6} \left( - 3 - 3 \frac{a \ln b - b \ln a}{a-b} - \frac{3}{2}  \frac{a \ln^2 b - b \ln^2 a}{a-b} + \frac{a \ln^3 b - b \ln^3 a}{a-b} \right) \right] + {\cal O}(\epsilon^3) \; , \vspace{0.25 cm} \\
\label{I3_div_1}
\end{equation}

\begin{equation*}
  - \frac{(1- \epsilon)}{\epsilon} \int_0^1 dx \frac{(A-B)^{\epsilon}}{A^{2}} = - \frac{c^{\epsilon}}{ab} \left \{ \frac{1}{\epsilon} -1 - \frac{a+b}{a-b} \ln \left( \frac{a}{b} \right)-\epsilon \left[ \zeta(2) + \frac{a+b}{a-b} \left( {\rm{Li}}_2 \left( 1- \frac{a}{b} \right) \right. \right. \right.
\end{equation*}
\begin{equation*}
    \left. \left. \left. - {\rm{Li}}_2 \left( 1- \frac{b}{a} \right) - \ln \left( \frac{a}{b} \right) \right) \right] + \epsilon^2 \left[ \zeta(2) - \frac{a+b}{a-b} \ln \left( \frac{a}{b} \right) \left( \zeta(2) + \ln \left( \frac{a}{b} \right) \left( -\frac{1}{2}- \ln \left( 1 - \frac{a}{b} \right) \right. \right. \right. \right. 
\end{equation*}
\begin{equation}
  \left. \left. \left. \left.  + \frac{1}{6} \ln \left( \frac{a}{b} \right) \right) \right) - \frac{4 b}{a-b} \zeta (3) + 2 \frac{a+b}{a-b} \left( {\rm{Li}}_2 \left( 1- \frac{a}{b} \right) + 2 {\rm{Li}}_3 \left( 1- \frac{a}{b} \right) + {\rm{Li}}_3 \left( \frac{a}{b} \right) \right) \right] \right \} \ + {\cal O}(\epsilon^3)  , \vspace{0.25 cm} \\
\label{I3_div_2}
\end{equation}

\begin{equation*}
    \int_0^1 dx \frac{1}{A(A-B)^{1-\epsilon}} = \frac{1}{abc} \left \{ \frac{a+b}{\epsilon} + (a+b) \ln c - (a-b) \ln \left( \frac{a}{b} \right) + \epsilon \left[ \frac{(a+b) \ln^2 c}{2}  \right. \right.
\end{equation*}
\begin{equation*}
   \left. - (a-b) \ln c \ln \left( \frac{a}{b} \right) -(a+b) \zeta (2) - (a-b) \left( {\rm{Li}}_2 \left( 1 - \frac{a}{b} \right) - {\rm{Li}}_2 \left( 1 - \frac{b}{a} \right) \right) \right ] + \epsilon^2 \left[ \frac{a+b}{6} \ln^3 c \right. 
\end{equation*}
\begin{equation*}
  \left. \left. \left. - \frac{a-b}{2} \ln \left( \frac{a}{b} \right) \ln^2 c - \ln c \bigg ( (a+b) \zeta(2) + (a-b) \left( {\rm{Li}}_2 \left( 1 - \frac{a}{b} \right) - {\rm{Li}}_2 \left( 1 - \frac{b}{a} \right) \right) \right) + 4 b \zeta (3) \right. \right.
\end{equation*}
\begin{equation*}
     \left. \left. \hspace{-0.15 cm} - (a-b) \left( \ln \left( \frac{a}{b} \right) \left( \zeta (2) -\ln \left( 1 - \frac{a}{b} \right) \ln \left( \frac{a}{b} \right) + \frac{1}{6} \ln^2 \left( \frac{a}{b} \right) \right) - 4 {\rm{Li}}_3 \left( 1 - \frac{a}{b} \right) - 2 {\rm{Li}}_3 \left( \frac{a}{b} \right)   \right)  \right] \right \} 
\end{equation*}
\begin{equation}
    + {\cal O}(\epsilon^3) \; ,
\label{I3_div_3}
\end{equation}
where $a= \vec{q}_1^{\; 2}$, $b= \vec{q}_2^{\; 2}$, $c=\vec{p}^{\; 2}$.
Plugging into (\ref{I3final}) the results given in (\ref{I3_div_1})-(\ref{I3_div_3}), one gets 
an expression for $\mathcal{I}_3$, valid up to the order $\epsilon^2$, in terms of finite 
one-dimensional integrals. A quick numerical comparison shows that this expression is perfectly equivalent to the one obtained in the previous Section. \\

By limiting the accuracy to the order $\epsilon$, we can express the result in a very compact form. We calculate the two residual one-dimensional integrals,
\begin{equation}
    \int_0^1 dx \frac{1}{A^2} \ln \left( \frac{A-B}{A} \right) = - \frac{1}{2ab} \left[ 2 - \ln \left( \frac{c^2}{a b} \right) + \frac{a+b}{a-b} \ln \left( \frac{a}{b} \right) \right] \; ,
\end{equation}
\begin{equation*}
   -\int_0^1 dx \frac{1}{A^2} {\rm Li}_2\left(\frac{B}{A}\right) = \frac{1}{2 a b} \left[ 2 -2 \zeta (2) - \ln \left( \frac{c^2}{ab} \right) + \frac{a+b}{a-b} \ln \left( \frac{a}{b} \right) + \frac{(a-b)^2 - c (a+b)}{c(a-b)} \right.
\end{equation*}
\begin{equation}
  \left.  \times \left( {\rm{Li}}_2 \left( 1 - \frac{a}{b} \right) - {\rm{Li}}_2 \left( 1 - \frac{b}{a} \right) + \frac{1}{2} \ln \left( \frac{c^2}{ab} \right) \ln \left( \frac{a}{b} \right) \right) \right] + \frac{(a-b)^2 -2 c(a+b) +c^2}{2 a b c} I_{a,b,c} \; ,
\end{equation}
where
\begin{equation}
    I_{a,b,c} = \int_0^1 dx \ \frac{1}{a x + b (1-x) - c x (1-x)} \ln \left( \frac{a x + b (1-x)}{c x(1-x)} \right) \; .
\label{FadinGorba}
\end{equation}
Various properties and representations of the integral~(\ref{FadinGorba}), together with its explicit value for $I_{\vec{q}_1^{\; 2}, \vec{q}_2^{\; 2}\vec{p}^{\; 2}}$, are given in Appendix~\ref{Appendix B}. Combining everything, we find
\begin{equation*}
    \mathcal{I}_3 = \frac{\Gamma^2 (1+\epsilon)}{\epsilon \Gamma (1+2 \epsilon)} \left[ (\vec{p}^{\; 2})^{\epsilon} \left( \frac{1}{\vec{p}^{ \; 2} \vec{q}_1^{\; 2}} + \frac{1}{\vec{p}^{\; 2} \vec{q}_2^{\; 2}} - \frac{1}{\vec{q}_1^{ \; 2} \vec{q}_2^{\; 2}} \right) \right.
\end{equation*}
\begin{equation*}
    \left. + (\vec{q}_1^{\; 2})^{\epsilon} \left( \frac{1}{ \vec{q}_1^{\; 2} \vec{q}_2^{\; 2}} + \frac{1}{\vec{q}_1^{\; 2} \vec{p}^{\; 2}} - \frac{1}{ \vec{q}_2^{\; 2} \vec{p}^{\; 2} } \right) + (\vec{q}_2^{\; 2})^{\epsilon} \left( \frac{1}{ \vec{q}_2^{\; 2} \vec{q}_1^{\; 2}} + \frac{1}{\vec{q}_2^{\; 2} \vec{p}^{\; 2}} - \frac{1}{ \vec{q}_1^{\; 2} \vec{p}^{\; 2} } \right) \right.
\end{equation*}
\begin{equation}
    + \frac{\epsilon^2}{\vec{q}_1^{\; 2} \vec{q}_2^{\; 2} \vec{p}^{\;2}} ((\vec{p}^{\;2})^2+(\vec{q}_1^{\; 2})^2+(\vec{q}_2^{\; 2})^2-2 \vec{q}_1^{\; 2} \vec{q}_2^{\; 2}-2 \vec{q}_1^{\; 2} \vec{p}^{\; 2} -2 \vec{q}_2^{\; 2} \vec{p}^{\; 2} ) I_{\vec{q}_1^{\; 2}, \vec{q}_2^{\; 2}, \vec{p}^{\; 2}} \; ,
\end{equation}
or 
\begin{equation*}
    \mathcal{I}_3 = \frac{\Gamma^2 (1+\epsilon)}{\epsilon \Gamma (1+2 \epsilon)}
\end{equation*}    
\begin{equation}
   \times \hat{\mathcal{S}} \left \{ (\vec{q}_2^{\; 2})^{\epsilon} \left( \frac{1}{ \vec{q}_2^{\; 2} \vec{q}_1^{\; 2}} + \frac{1}{\vec{q}_2^{\; 2} \vec{p}^{\; 2}} - \frac{1}{ \vec{q}_1^{\; 2} \vec{p}^{\; 2} } \right)
    + \frac{\epsilon^2}{\vec{q}_1^{\; 2} \vec{q}_2^{\; 2} \vec{p}^{\; 2}} ((\vec{q}_2^{\; 2})^2- \vec{q}_1^{\; 2} \vec{q}_2^{\; 2}- \vec{q}_2^{\; 2} \vec{p}^{\; 2} ) I_{\vec{q}_1^{\; 2}, \vec{q}_2^{\; 2}, \vec{p}^{\; 2}} \right \} \; .
\end{equation}
This result, after multiplication by the factor $\pi^{1+\epsilon} \Gamma(1-\epsilon)$ (see the definition~(\ref{I3prime_def})), is equivalent to (\ref{I3BDKe^2}) at the order $\epsilon$.

\subsection{$\mathcal{I}_{4 B}$ and $\mathcal{I}_{4 A}$: BDK method}
\label{subsec:I41}
The integral that has to be evaluated is
\begin{equation*}
\mathcal{I}_{4B} = \int_0^1 \frac{dx}{x} \int \frac{d^{D-2} k}{\pi^{1 + \epsilon} \Gamma (1-\epsilon)} 
\end{equation*}
\begin{equation}
   \times \left[ \frac{1-x}{\left( x \vec{k}^2 + (1-x) (\vec{k}-\vec{q}_1)^2 \right) (\vec{k}-(1-x)(\vec{q}_1-\vec{q}_2))^2} - \frac{1}{(\vec{k}-\vec{q}_{1})^2 (\vec{k}-(\vec{q}_1-\vec{q}_2))^2} \right] \; ,
\tag{\ref{I4Bcal}}
\end{equation}
Let us start from the integral $I_{4B}$ defined as 
\begin{equation}
  I_{4B} = \frac{1}{i} \int d^{D}k \frac{1}{(k^{2} + i \varepsilon) [(k+q_1)^{2}+i\varepsilon]
    [(k+q_2)^{2}+i\varepsilon] [(k-p_B)^2 + i \varepsilon]}\;.
\label{I4B}
\end{equation}
Using the BDK method and working in the Euclidean region ($s,s_1,s_2,t_1,t_2<0$) the result for this integral is~\cite{Bern_1994}
\begin{equation}
\begin{split}
I_{4B} = & \frac{\pi^{2+\epsilon}}{s_2 t_2} \frac{\Gamma(1-\epsilon) \Gamma^2(1+\epsilon)}{\Gamma(1+2 \epsilon)} \frac{2}{\epsilon^2} \left[ (-\alpha_4 (\alpha_1 - \alpha_5))^{-\epsilon} \; _2F_1 \left( \epsilon, \epsilon, 1+\epsilon; \frac{\alpha_1 \alpha_4 + \alpha_5 \alpha_3 - \alpha_5 \alpha_4}{\alpha_4 (\alpha_1 - \alpha_5)} \right) \right. \\ & \left. +(-\alpha_5 (\alpha_3 - \alpha_4))^{-\epsilon} \; _2F_1 \left( \epsilon, \epsilon, 1+\epsilon; \frac{\alpha_1 \alpha_4 + \alpha_5 \alpha_3 - \alpha_5 \alpha_4}{\alpha_5 (\alpha_3 - \alpha_4)} \right) \right. \\ & \left. -((\alpha_1-\alpha_5) (\alpha_3 - \alpha_4))^{-\epsilon} \; _2F_1 \left( \epsilon, \epsilon, 1+\epsilon; -\frac{\alpha_1 \alpha_4 + \alpha_5 \alpha_3 - \alpha_5 \alpha_4}{(\alpha_1-\alpha_5) (\alpha_3 - \alpha_4)} \right) \right]\;,
\end{split}
\label{C}
\end{equation}
where
\begin{equation*}
  \alpha_1 = \sqrt{-\frac{s_1 s_2}{s t_2 t_1}}\;, \hspace{0.5 cm}
  \alpha_2 = \sqrt{-\frac{s_2 t_2}{s s_1 t_1}}\;, \hspace{0.5 cm}
  \alpha_3 = \sqrt{-\frac{s t_2}{s_2 s_1 t_1}}\;, 
\end{equation*}
\begin{equation}
  \alpha_4 = \sqrt{-\frac{s t_1}{s_1 s_2 t_2}}\;, \hspace{0.5 cm}
  \alpha_5 = \sqrt{-\frac{s_1 t_1}{s s_2 t_2}} \;.
  \label{alpha}
\end{equation}
We then have
\begin{equation}
\begin{split}
I_{4B} = & \frac{\pi^{2+\epsilon}}{s_2 t_2} \frac{\Gamma(1-\epsilon) \Gamma^2(1+\epsilon)}{\Gamma(1+2 \epsilon)} \frac{2}{\epsilon^2} \left[ \frac{(-s_2)^{\epsilon} (-t_2)^{\epsilon}}{(s_2-t_1)^{\epsilon}} \; _2F_1 \left( \epsilon, \epsilon, 1+\epsilon; 1 - \frac{(-t_2)}{s_2-t_1} \right) \right. \\ & \left. +\frac{(-s_2)^{\epsilon} (-t_2)^{\epsilon}}{(t_2-t_1)^{\epsilon}} \; _2F_1 \left( \epsilon, \epsilon, 1+\epsilon; 1 - \frac{s_2}{t_1-t_2} \right) \right. \\ & \left. - \frac{(-s_2)^{\epsilon} (-t_2)^{\epsilon} (-t_1)^{\epsilon}}{(s_2-t_1)^{\epsilon} (t_2-t_1)^{\epsilon}} \; _2F_1 \left( \epsilon, \epsilon, 1+\epsilon; 1 - \frac{s_2 t_2}{(s_2-t_1)(t_2-t_1)} \right)  \right]\; .
\end{split}
\label{C1}
\end{equation} 
Using~(\ref{HyperProp1}), this result can be put in the following simpler form:
\begin{equation} 
I_{4B} = \frac{\pi^{2+\epsilon}}{s_2 t_2} \frac{\Gamma(1-\epsilon) \Gamma^2(1+\epsilon)}{\Gamma(1+2 \epsilon)} \frac{2}{\epsilon^2} \left[ (-s_2)^{\epsilon}\; _2F_1 \left(1, \epsilon, 1+\epsilon; 1 - \frac{s_2-t_1}{(-t_2)} \right) \right.
\label{C1bis}
\end{equation} 
\[
\left. + (-t_2)^{\epsilon} \; _2F_1 \left(1, \epsilon, 1+\epsilon; 1 - \frac{t_1-t_2}{s_2} \right)
- (-t_1)^{\epsilon} 
\; _2F_1 \left(1, \epsilon, 1+\epsilon; 1 - \frac{(s_2-t_1)(t_1-t_2)}{s_2(-t_2)} \right)  \right]\; .
\]
We re-derive this result by a different method in Appendix~\ref{Appendix B}. \\
For the first term in the square roots in~(\ref{C}) we have
\begin{equation}
\left(-t_2\right)^\epsilon\left(-s_2\right)^\epsilon\left(s_2-t_1\right)^{-\epsilon}{ }_2 F_1\left(\epsilon, \epsilon ; 1+\epsilon ; 1- \frac{\left(-t_2\right)}{s_2-t_1}\right) \; .
\end{equation}
Analytically continuing the result in the region of positive $s_2$ by the replacement
\begin{equation}
    (-s_2) \rightarrow e^{-i \pi } s_2 \;,
\end{equation}
using the MRK approximation to neglect $(-t_2)/(s_2-t_1)$ with respect to one
in the argument of the hypergeometric function, and recalling that
\begin{equation}
    { }_2 F_1(\epsilon, \epsilon , 1+\epsilon ; 1)=\Gamma(1-\epsilon) \Gamma(1+\epsilon)=\frac{\pi \epsilon}{\sin (\pi \epsilon)} \; ,
\end{equation}
we obtain
\begin{equation}
\left(-t_2\right)^\epsilon e^{-i \pi \epsilon} \Gamma(1-\epsilon) \Gamma(1+\epsilon)=\left(-t_2\right)^\epsilon\left[\cos (\pi \epsilon) \frac{\pi \epsilon}{\sin (\pi \epsilon)}-i \pi \epsilon\right] \; .
\end{equation}
The second term,
\begin{equation}
    \left(\frac{-t_2 s_2}{t_1-t_2} \right)^\epsilon{ }_2 F_1\left(\epsilon, \epsilon , 1+\epsilon ; 1- \frac{s_2}{t_1-t_2} \right) \;,
\end{equation}
with the help of~(\ref{HyperProp1}) and~(\ref{HyperExp2}) gives, assuming $t_1 > t_2$, 
\begin{equation}
    \left(-t_2\right)^\epsilon{ }_2 F_1\left(1, \epsilon , 1+\epsilon ; 1 - \frac{ t_1-t_2 }{s_2} \right)=\left(-t_2\right)^\epsilon\left(1-\epsilon \ln \left( \frac{t_1-t_2 }{s_2} \right)-\sum_{n=2}^{\infty}(-\epsilon)^n \zeta(n) \right ) \; .
\end{equation}
The third term,
\begin{equation}
    -\left( \frac{t_1 t_2}{t_1-t_2} \right)^\epsilon{ }_2 F_1\left(\epsilon, \epsilon , 1+\epsilon ; 1+ \frac{t_2}{t_1-t_2} \right) \;,
\end{equation}
using~(\ref{HyperProp1}) and~(\ref{HyperExp2}), again assuming $t_1 > t_2$, becomes
\begin{equation}
  -\left(-t_1\right)^\epsilon{ }_2 F_1\left(1, \epsilon , 1+\epsilon ; \frac{t_1}{t_2} \right)=-\left(-t_1\right)^\epsilon\left(1-\epsilon \ln \left(1- \frac{t_1}{t_2} \right) - \sum_{n=2}^{\infty}(-\epsilon)^n {\rm{Li}}_n\left( \frac{t_1}{ t_2} \right)\right) \; .  
\end{equation}
We finally obtain
\begin{gather}
    I_{4B} = \frac{\pi^{2+\epsilon}}{s_2 t_2} \frac{\Gamma(1-\epsilon) \Gamma^2(1+\epsilon)}{\Gamma(1+2 \epsilon)} \frac{2}{\epsilon^2} \left(-t_1\right)^\epsilon \left[ \left(\frac{t_2}{t_1}\right)^\epsilon \bigg ( e^{-i \pi \epsilon} \Gamma(1-\epsilon) \Gamma(1+\epsilon) \right. \nonumber \\
    \left. \left. + 1  -\epsilon \ln \left( \frac{ t_1-t_2 }{s_2} \right) -\sum_{n=2}^{\infty}(-\epsilon)^n \zeta(n) \right ) - 1 + \epsilon \ln \left(1- \frac{t_1}{t_2} \right) + \sum_{n=2}^{\infty}(-\epsilon)^n {\rm{Li}}_n\left( \frac{t_1}{ t_2} \right) \right ] \; .
    \label{I4B_BDK}
\end{gather}
Furthermore, starting from the integral in eq.~(\ref{I4B}), using the standard Sudakov decomposition for the 4-momentum $k$, one finds that, in the multi-Regge kinematics (see Appendix~\ref{Appendix C}),
\begin{equation}
I_{4B}=- \frac{\pi^{2+\epsilon} \Gamma (1-\epsilon)}{s_2} \left[ \frac{\Gamma^2 (\epsilon)}{\Gamma (2 \epsilon)} (-t_2)^{\epsilon -1} \left( \ln \left( \frac{- s_2}{- t_2} \right) + \psi (1-\epsilon) - 2 \psi (\epsilon) + \psi (2 \epsilon)  \right) + \mathcal{I}_{4B} \right] \;,
\label{I4B-I4Bcal}
\end{equation}
where $\mathcal{I}_{4B}$ is exactly the integral defined in eq.~(\ref{I4Bcal}). From this relation we can hence derive an expression for $\mathcal{I}_{4B}$ that, after some manipulations, can be cast in the following form:
\begin{gather}
    \mathcal{I}_{4 B}=\frac{\Gamma^2(1+\epsilon)}{\Gamma(1+2 \epsilon)} \frac{2}{\epsilon^2} \frac{\left(-t_1\right)^\epsilon}{-t_2}\left[\left(\frac{t_2}{t_1}\right)^\epsilon\left(-\frac{1}{2}+\frac{\pi \epsilon}{\sin (\pi \epsilon)} \cos (\pi \epsilon)-\epsilon \ln \left(1- \frac{t_1}{t_2} \right)\right.\right. \nonumber \\ 
    \left.\left.+\sum_{n=2}^{\infty} \epsilon^n \zeta(n)\left(1-(-1)^n\left(2^{n-1}-1\right)\right)\right)-1+\epsilon \ln \left(1- \frac{t_1}{t_2} \right)+\sum_{n=2}^{\infty}(-\epsilon)^n {\rm{Li}}_n\left( \frac{t_1}{t_2} \right)\right] \; .
\label{I4BcalFin}
\end{gather} 
The result remains valid for $t_2 > t_1$. In particular, truncating the summation at $n=4$ gives the desired result. \\
This result is completely equivalent to the one obtained by calculating directly $\mathcal{I}_{4B}$ introducing Feynman parameters (see the next subsection). The integral $\mathcal{I}_{4A}$ is defined similarly to $\mathcal{I}_{4B}$ in eq.~(\ref{I4Bcal}), up to the replacement $q_1 \rightarrow -q_2$, implying that
\begin{equation}
    \mathcal{I}_{4A} = \mathcal{I}_{4B} (t_1 \leftrightarrow t_2) \; .
\end{equation}

\subsection{$\mathcal{I}_{4 B}$ and $\mathcal{I}_{4 A}$: Alternative calculation}
\label{subsec:I42}
Starting from the expression
\begin{equation}
\mathcal{I}_{4B} = \int_0^1 \frac{dx}{x} \int \frac{d^{D-2} k}{\pi^{1 + \epsilon} \Gamma (1-\epsilon)} 
\tag{\ref{I4Bcal}}
\end{equation} 
\begin{equation*}
   \times \left[ \frac{1-x}{\left( x \vec{k}^2 + (1-x) (\vec{k}-\vec{q}_1)^2 \right) (\vec{k}-(1-x)(\vec{q}_1-\vec{q}_2))^2} - \frac{1}{(\vec{k}-\vec{q}_{1})^2 (\vec{k}-(\vec{q}_1-\vec{q}_2))^2} \right] \; ,
\end{equation*}
introducing the Feynman parametrization and performing the integration over $k$, we obtain
\begin{gather}
\mathcal{I}_{4 B}=\int_0^1 \frac{d x}{x}\left[ \int_0^1 d z \frac{(1-x)}{[z(1-x)(a x+b(1-x)(1-z))]^{1-\epsilon}}-\int_0^1 d z \frac{1}{[bz(1-z)]^{1-\epsilon}}\right] \; . 
\end{gather}
Defining
\begin{equation}
  a = \vec{q}_1^{\; 2} = - t_1 \; , \hspace{0.5 cm} b = \vec{q}_2^{\; 2} = - t_2 \; , \hspace{0.5 cm} c = 1 - \frac{a}{b} \; ,
\end{equation}
we can write $\mathcal{I}_{4 B} \equiv b^{\epsilon-1} F(c)$, with 
\begin{equation}
    F(c) = \int_0^1 \frac{dx}{x} \left[ \int_0^1 \frac{dz}{z} \frac{z^{\epsilon} (1-x)^{\epsilon}}{ (1-x c- z(1-x) 
 )^{1-\epsilon}} - \int_0^1 dz \; z^{\epsilon-1} (1-z)^{\epsilon-1} \right] \; . 
\end{equation}
Performing the transformation $y=(1-x)z$ in the first term, we get
\begin{equation}
    F (c) = \int_0^1 \frac{dx}{x} \left[ \int_0^{1-x} \frac{d y}{y} \frac{y^{\epsilon}}{ (1-x c- y 
 )^{1-\epsilon}} - \int_0^1 dz \; z^{\epsilon-1} (1-z)^{\epsilon-1} \right] \; .
\end{equation}
It is very simple to compute the integral for $c=0$; we find
\begin{gather}
    F (0) = \int_0^1 \frac{dx}{x} \left[ \int_0^{1-x} d y \; y^{\epsilon-1} (1-y)^{\epsilon-1} - \int_0^1 dz \; z^{\epsilon-1} (1-z)^{\epsilon-1} \right] \nonumber \\
    = - \int_0^1 \frac{dx}{x} \int_{1-x}^1 dz \; z^{\epsilon-1} (1-z)^{\epsilon-1} = - \int_{0}^1 dz \; z^{\epsilon-1} (1-z)^{\epsilon-1} \int_{1-z}^1 \frac{dx}{x} \nonumber \\ \hspace{1.0 cm} = \int_{0}^1 dz \; z^{\epsilon-1} (1-z)^{\epsilon-1} \ln (1-z) = \frac{d}{d \delta} \left[ \int_{0}^1 dz \; z^{\epsilon-1} (1-z)^{\delta-1} \right]_{\delta=\epsilon}  \; .
\end{gather}
The function $F(c)$ in $c=0$ is then
\begin{equation}
    F(0) = \frac{\Gamma^2 (\epsilon)}{\Gamma (2 \epsilon)} \left( \psi (\epsilon) - \psi (2 \epsilon) \right) \; .
\end{equation}
We now compute the derivative with respect to $c$ of the function $F$ and get
\begin{gather}
F^{\prime}(c)=(1-\epsilon) \int_0^1 dx \int_0^{1-x} \frac{d y}{y} y^\epsilon(1-x c-y)^{\epsilon-2} = (1-\epsilon) \int_0^1 \frac{d y}{y} y^\epsilon \int_0^{1-y} \hspace{-0.2 cm} d x (1-x c-y)^{\epsilon-2} \nonumber \\ =\frac{1}{c} \int_0^1 \frac{d y}{y} y^\epsilon\left(((1-c)(1-y))^{\epsilon-1}-(1-y)^{\epsilon-1}\right) =\frac{1}{c} \frac{\Gamma^2(\epsilon)}{\Gamma(2 \epsilon)}\left((1-c)^{\epsilon-1}-1\right) \; .
\end{gather}
Having this information, we can write
\begin{equation}
    F (c) = F (0) + \int_0^c dx \; F'(x) = \frac{ \Gamma^2 (\epsilon)}{\Gamma (2 \epsilon)} \left[ \psi (\epsilon) - \psi (2 \epsilon) + \int_0^c dx \left( \frac{(1-x)^{\epsilon-1} -1}{x} \right) \right] \;.
\end{equation}
The integral on the right-hand side can be computed to all orders in $\epsilon$:
\begin{gather}
    \int_0^c dx \left( \frac{(1-x)^{\epsilon-1} -1}{x} \right) = \int_0^c dx \left[ (1-x)^{\epsilon} \left( \frac{1}{1-x} + \frac{1}{x} \right) - \frac{1}{x} \right] \nonumber \\ = \frac{1}{\epsilon} \left[ 1- (1-c)^{\epsilon}  \right] + \int_0^c \frac{dx}{x} ((1-x)^{\epsilon}-1) = \sum_{n=1}^{\infty} \left( \frac{\epsilon^n}{n!} \int_0^1 \frac{dx}{x} \ln^n (1-cx) -\epsilon^{n-1} \frac{\ln^n (1-c)}{n!}  \right).
\end{gather}
Using
\begin{equation}
    \frac{1}{n!} \int_0^1 \frac{dx}{x} \ln^n (1-c x) = (-1)^n \;  {\rm{S}}_{1,n} (c) \; ,
\end{equation}
we finally find
\begin{equation}
    \int_0^c dx \left( \frac{(1-x)^{\epsilon-1} -1}{x} \right) = \sum_{n=1}^{\infty} \epsilon^{n-1} \left( -\frac{\ln^n (1-c)}{n!} + \epsilon \; (-1)^n \;  {\rm{S}}_{1,n} (c) \right) \; .
\end{equation}
The final result for $\mathcal{I}_{4B}$ is then 
\begin{equation}
    \mathcal{I}_{4 B} = \frac{ \Gamma^2 (\epsilon)}{\Gamma (2 \epsilon)} (-t_2)^{\epsilon-1} \left[ \psi (\epsilon) - \psi (2 \epsilon) + \sum_{n=1}^{\infty} \epsilon^{n-1} \left( -\frac{\ln^n \left( t_1 / t_2 \right)}{n!} + \epsilon \; (-1)^n \;  {\rm{S}}_{1,n} \left( 1 - \frac{t_1}{t_2} \right) \right) \right] \; .
    \label{CalI4BDerTrick}
\end{equation}
The expression~(\ref{CalI4BDerTrick}) gives an alternative representation of $\mathcal{I}_{4 B}$, which is completely equivalent to (\ref{I4BcalFin}). Again, the complete $\epsilon^2$ result is obtained truncating the series at $n=4$. \\
We can achieve an alternative representation in terms of the hypergeometric function. In fact, we note that
\begin{gather} 
\int_0^c \frac{d x}{x}\left((1-x)^\epsilon-1\right)=\left((1-c)^\epsilon-1\right) \ln c+\epsilon\left(\int_0^1 \frac{d y}{y} \ln (1-y) y^\epsilon-\int_0^{1-c} \frac{d y}{y} \ln (1-y) y^\epsilon\right) \nonumber \\ =\left((1-c)^\epsilon-1\right) \ln c+\psi(1)-\psi(1+\epsilon)-\frac{(1-c)^\epsilon}{\epsilon}\bigg( { }_2 F_1(1, \epsilon, 1+\epsilon ; 1-c)-1+\epsilon \ln c \bigg) .
\label{HyperReprIntInCal4}
\end{gather}
The last equality can be proved by using the explicit result for the first integral, \textit{i.e.} 
\begin{equation} 
\int_0^1 \frac{d y}{y} \ln (1-y) y^{\epsilon} = \left. \frac{d}{d \delta} \int_0^1 \frac{d y}{y}(1-y)^\delta y^\epsilon\right|_{\delta=0} = \left. \frac{d}{d \delta} \frac{\Gamma(1+\delta) \Gamma(\epsilon)}{\Gamma(1+\delta+\epsilon)}\right|_{\delta=0}= \frac{1}{\epsilon}(\psi(1)-\psi(1+\epsilon)) \; ,
\end{equation}
and the following expression for the second, in terms of the hypergeometric function,
\begin{gather}
{ }_2 F_1(1, \epsilon, 1+\epsilon ; 1-c)=\epsilon \int_0^1 \frac{d y}{y(1-y(1-c))} y^\epsilon=\epsilon \int_0^1 y^\epsilon d \ln \frac{y}{(1-y(1-c))} \nonumber \\ =-\epsilon \ln c -\epsilon^2 \int_0^1 d y y^{\epsilon-1} \ln y+\epsilon^2 \int_0^1 d y y^{\epsilon-1} \ln (1-y(1-c)) \nonumber \\  = -\epsilon \ln c+1+\epsilon^2(1-c)^{-\epsilon} \int_0^{1-c} d y y^{\epsilon-1} \ln (1-y) \;. 
\end{gather}
Using~(\ref{HyperReprIntInCal4}), we obtain
\begin{equation}
    \mathcal{I}_{4 B}=\frac{\Gamma^2(\epsilon)}{\Gamma(2 \epsilon)} b^{\epsilon-1}\left(\frac{1}{2 \epsilon}+\psi(1)-\psi(1+2 \epsilon) - \ln c -\frac{(1-c)^\epsilon}{\epsilon}{ }_2 F_1(1, \epsilon, 1+\epsilon ; 1-c)\right) \; ,
\end{equation}
which is completely equivalent to both (\ref{I4BcalFin}) and (\ref{CalI4BDerTrick}). 
\subsection{Combination $\mathcal{I}_{5} - \mathcal{L}_3$}
\label{subsec:I5I3}
Let us start by considering the pentagonal integral $I_5$, defined as
\begin{equation}
  I_{5} = \frac{1}{i} \int d^{D}k \frac{1}{(k^{2} + i \varepsilon) [(k+q_1)^{2}+i\varepsilon]
    [(k+q_2)^{2}+i\varepsilon] [(k+p_A)^2 + i \varepsilon]
    [(k-p_B)^2 + i \varepsilon]}
  \;.
  \label{I5}
\end{equation}
In multi-Regge kinematics (and in $D=4+2 \epsilon$), it is given by the following combination:
\begin{equation}
I_5 = \frac{\pi^{2+\epsilon} \Gamma (1-\epsilon)}{s} \left[ \ln \left( \frac{(-s) (\vec{q}_1-\vec{q}_2)^2}{(-s_1)(-s_2)} \right) \mathcal{I}_3 + \mathcal{L}_3 -\mathcal{I}_5 \right]
\label{I5inTransver}
\end{equation}
Inverting this relation, we obtain
\begin{equation}
 \mathcal{I}_5 - \mathcal{L}_3 = \ln \left( \frac{(-s) (\vec{q}_1-\vec{q}_2)^2}{(-s_1)(-s_2)} \right) \mathcal{I}_3 - \frac{s}{\pi^{2+\epsilon} \Gamma (1-\epsilon)} I_5 \; .
\end{equation}
Hence, the integral $I_{5}$ can be used (together with $\mathcal{I}_{3}$) to calculate
$\mathcal{I}_{5} - \mathcal{L}_3$ (this combination is the one appearing in the Lipatov vertex~\cite{Fadin:2000yp}). \\
We start by working in the Euclidean region, defined by $s,s_1,s_2,t_1,t_2 <0$, the analytical continuation to the physical region being obtained according to the prescriptions
\begin{equation}
    (-s) \rightarrow e^{-i \pi } s \; , \hspace{1 cm} (-s_1) \rightarrow e^{-i \pi } s_1 \; , \hspace{1 cm} (-s_2) \rightarrow e^{-i \pi} s_2 \; .
\end{equation}
We use the quantities $\alpha_i$ defined before, which we recall here:
\begin{equation*}
  \alpha_1 = \sqrt{-\frac{s_1 s_2}{s t_2 t_1}}\;, \hspace{0.5 cm}
  \alpha_2 = \sqrt{-\frac{s_2 t_2}{s s_1 t_1}}\;, \hspace{0.5 cm}
  \alpha_3 = \sqrt{-\frac{s t_2}{s_2 s_1 t_1}}\;, 
\end{equation*}
\begin{equation}
  \alpha_4 = \sqrt{-\frac{s t_1}{s_1 s_2 t_2}}\;, \hspace{0.5 cm}
  \alpha_5 = \sqrt{-\frac{s_1 t_1}{s s_2 t_2}} \;.
\tag{\ref{alpha}}
\end{equation}
Let us define the reduced integral $\hat{I}_5$ in this way:
\begin{equation}
I_5= -\pi^{D/2} \alpha_1 \alpha_2 \alpha_3 \alpha_4 \alpha_5 \hat{I}_5 \;.
\end{equation}
In Ref.~\cite{Bern_1994} it is shown that this integral is given by the
recursive relation
\begin{equation}
  \hat{I}_5 = \frac{1}{2} \left[ \sum_{i=1}^{5} \gamma_i \hat{I}_4^{(i)} -
    2 \epsilon \Delta_5 \hat{I}_5^{(D = 6+2 \epsilon)} \right]\;,
    \label{IterativeRelI5}
\end{equation}
where $\Delta_5$ is the following quantity:
\begin{equation}
  \Delta_5 = \sum_{i=1}^5 (\alpha_i^2 - 2 \alpha_i \alpha_{i+1}
  + 2 \alpha_i \alpha_{i+2}) \;.
\end{equation} 
The $\gamma_i$'s are
\begin{eqnarray}
\gamma_1 &=& \alpha_1 -\alpha_2 + \alpha_3 + \alpha_4 - \alpha_5 \;, \\
\gamma_2 &=& -\alpha_1 +\alpha_2 - \alpha_3 + \alpha_4 + \alpha_5 \;, \\
\gamma_3 &=& \alpha_1 - \alpha_2 + \alpha_3 - \alpha_4 + \alpha_5 \;, \\
\gamma_4 &=& \alpha_1 + \alpha_2 - \alpha_3 + \alpha_4 - \alpha_5 \;, \\
\gamma_5 &=& -\alpha_1 + \alpha_2 + \alpha_3 - \alpha_4 + \alpha_5 \;. 
\end{eqnarray}
$\hat{I}_4^{(i)}$ are the reduced version of $I_4^{(i)}$ (apart for a trivial $\pi^{2+\epsilon}$), \textit{i.e.} 
\begin{equation}
    \hat{I}_4^{(i)} = \frac{\alpha_1 \alpha_2 \alpha_3 \alpha_4 \alpha_5}{\alpha_i} I_4^{(i)} \;.
\end{equation}
\subsubsection{Box integrals part}
\label{subsubsec:Boxes}
\begin{figure}
\begin{picture}(400,120)
\put(145,0){\includegraphics[width=0.4\textwidth]{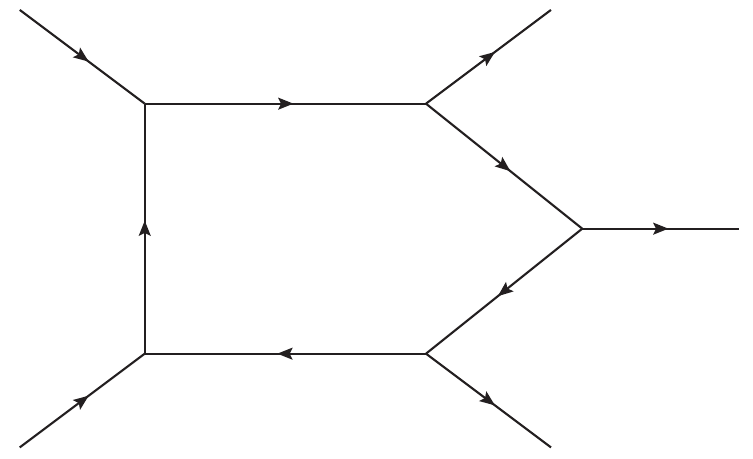}}
\put(165,50){$k$}
\put(170,95){$p_A$}
\put(215,95){$p_A-q_1$}
\put(170,09){$p_B$}
\put(215,09){$p_B+q_2$}
\put(283,65){$q_1-q_2$}
\end{picture}
\caption{Schematic representation of the pentagonal integral. The arrows denote the direction of the momenta.}
\label{Fig:Pentagon}
\end{figure}
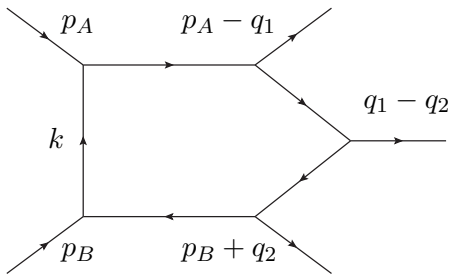
The integrals $I_4^{(i)}$ are box integrals with all external legs on-shell,
but one, which is off-shell. They can be obtained starting from $I_5$ (see Fig.~\ref{Fig:Pentagon}) in this
way (apart from a factor $\pi^{2+\epsilon}$):
\begin{itemize}
\item $I_4^{(1)}$ \\
  Making the propagator between $p_A$ and $p_B$ ``collapse'', so that the two
  on-shell legs associated with $p_A$ and $p_B$ become a unique off-shell
  external leg with (incoming) momenta $p_A+p_B$.
\item $I_4^{(2)}$ \\
  Making the propagator between $p_A$ and $p_{A'} = p_A - q_1$ ``collapse'',
  so that the two on-shell legs associated with $p_A$ and $p_{A'}$ become a
  unique off-shell external leg with (incoming) momenta $q_1$.    
\item $I_4^{(3)}$ \\
  Making the propagator between $p_{A'} = p_A - q_1$ and $q_1 - q_2$
  ``collapse'', so that the two on-shell legs associated with $p_{A'}$ and
  $q_1 - q_2$ become a unique off-shell external leg with (outgoing) momenta
  $p_A - q_2$.   
\item $I_4^{(4)}$ \\
  Making the propagator between $q_1-q_2$ and $p_{B'} = p_B + q_2$ ``collapse'',
  so that the two on-shell legs associated with $q_1-q_2$ and $p_{B'}$ become
  a unique off-shell external leg with (outgoing) momenta $p_B + q_1$.   
\item $I_4^{(5)}$ \\
  Making the propagator between $p_{B}$ and $p_{B'}$ ``collapse'', so that the
  two on-shell legs associated with $p_{B}$ and $p_{B'}$ become a unique
  off-shell external leg with (outgoing) momenta $q_2$. 
\end{itemize} 
The results for the reduced version of these integrals are
\begin{equation}
\begin{split}
\hat{I}_{4}^{(1)} (s_1,s_2,s) = & \frac{2}{\epsilon^2} r_{\Gamma} \left[ (-\alpha_3 (\alpha_5 - \alpha_4))^{-\epsilon} \; _2F_1 \left( \epsilon, \epsilon, 1+\epsilon; \frac{\alpha_5 \alpha_3 + \alpha_2 \alpha_4 - \alpha_4 \alpha_3}{\alpha_3 (\alpha_5 - \alpha_4)} \right) \right. \\ & \left. +(-\alpha_4 (\alpha_2 - \alpha_3))^{-\epsilon} \; _2F_1 \left( \epsilon, \epsilon, 1+\epsilon; \frac{\alpha_5 \alpha_3 + \alpha_2 \alpha_4 - \alpha_4 \alpha_3}{\alpha_4 (\alpha_2 - \alpha_3)} \right) \right. \\ & \left. -((\alpha_5-\alpha_4) (\alpha_2 - \alpha_3))^{-\epsilon} \; _2F_1 \left( \epsilon, \epsilon, 1+\epsilon; -\frac{\alpha_5 \alpha_3 + \alpha_2 \alpha_4 - \alpha_4 \alpha_3}{(\alpha_5-\alpha_4) (\alpha_2 - \alpha_3)} \right) \right] \;,
\end{split}
\end{equation}
\begin{equation}
\begin{split}
\hat{I}_{4}^{(2)} (s_2,t_2,t_1) = & \frac{2}{\epsilon^2} r_{\Gamma} \left[ (-\alpha_4 (\alpha_1 - \alpha_5))^{-\epsilon} \; _2F_1 \left( \epsilon, \epsilon, 1+\epsilon; \frac{\alpha_1 \alpha_4 + \alpha_3 \alpha_5 - \alpha_4 \alpha_5}{\alpha_4 (\alpha_1 - \alpha_5)} \right) \right. \\ & \left. + (-\alpha_5 (\alpha_3 - \alpha_4))^{-\epsilon} \; _2F_1 \left( \epsilon, \epsilon, 1+\epsilon; \frac{\alpha_1 \alpha_4 + \alpha_5 \alpha_3 - \alpha_5 \alpha_4}{\alpha_5 (\alpha_3 - \alpha_4)} \right) \right. \\ & \left. -((\alpha_1-\alpha_5) (\alpha_3 - \alpha_4))^{-\epsilon} \; _2F_1 \left( \epsilon, \epsilon, 1+\epsilon; -\frac{\alpha_1 \alpha_4 + \alpha_5 \alpha_3 - \alpha_5 \alpha_4}{(\alpha_1-\alpha_5) (\alpha_3 - \alpha_4)} \right) \right] \; ,
\end{split}
\end{equation}
\begin{equation}
\begin{split}
\hat{I}_{4}^{(3)} (s,s_1,t_2) = &  \frac{2}{\epsilon^2} r_{\Gamma} \left[ (-\alpha_5 (\alpha_2 - \alpha_1))^{-\epsilon} \; _2F_1 \left( \epsilon, \epsilon, 1+\epsilon; \frac{\alpha_2 \alpha_5 + \alpha_1 \alpha_4 - \alpha_1 \alpha_5}{\alpha_5 (\alpha_2 - \alpha_1)} \right) \right. \\ & \left. +(-\alpha_1 (\alpha_4 - \alpha_5))^{-\epsilon} \; _2F_1 \left( \epsilon, \epsilon, 1+\epsilon; \frac{\alpha_2 \alpha_5 + \alpha_1 \alpha_4 - \alpha_1 \alpha_5}{\alpha_1 (\alpha_4 - \alpha_5)} \right) \right. \\ & \left. -((\alpha_2-\alpha_1) (\alpha_4 - \alpha_5))^{-\epsilon} \; _2F_1 \left( \epsilon, \epsilon, 1+\epsilon; -\frac{\alpha_2 \alpha_5 + \alpha_1 \alpha_4 - \alpha_1 \alpha_5}{(\alpha_2-\alpha_1) (\alpha_4 - \alpha_5)} \right) \right]\;,
\end{split}
\end{equation}
\begin{equation}
\begin{split}
\hat{I}_{4}^{(4)} (s,s_2,t_1) = & \frac{2}{\epsilon^2} r_{\Gamma} \left[ (-\alpha_1 (\alpha_3 - \alpha_2))^{-\epsilon} \; _2F_1 \left( \epsilon, \epsilon, 1+\epsilon; \frac{\alpha_1 \alpha_3 + \alpha_2 \alpha_5 - \alpha_2 \alpha_1}{\alpha_1 (\alpha_3 - \alpha_2)} \right) \right. \\ & \left. +(-\alpha_2 (\alpha_5 - \alpha_1))^{-\epsilon} \; _2F_1 \left( \epsilon, \epsilon, 1+\epsilon; \frac{\alpha_1 \alpha_3 + \alpha_2 \alpha_5 - \alpha_2 \alpha_1}{\alpha_2 (\alpha_5 - \alpha_1)} \right) \right. \\ & \left. -((\alpha_3-\alpha_2) (\alpha_5 - \alpha_1))^{-\epsilon} \; _2F_1 \left( \epsilon, \epsilon, 1+\epsilon; -\frac{\alpha_1 \alpha_3 + \alpha_2 \alpha_5 - \alpha_2 \alpha_1}{(\alpha_3-\alpha_2) (\alpha_5 - \alpha_1)} \right) \right]\;,
\end{split}
\end{equation}
\begin{equation}
\begin{split}
\hat{I}_{4}^{(5)} (s_1,t_1,t_2) = & \frac{2}{\epsilon^2} r_{\Gamma} \left[ (-\alpha_3 (\alpha_1 - \alpha_2))^{-\epsilon} \; _2F_1 \left( \epsilon, \epsilon, 1+\epsilon; \frac{\alpha_1 \alpha_3 + \alpha_2 \alpha_4 - \alpha_2 \alpha_3}{\alpha_3 (\alpha_1 - \alpha_2)} \right) \right. \\ & \left. +(-\alpha_2 (\alpha_4 - \alpha_3))^{-\epsilon} \; _2F_1 \left( \epsilon, \epsilon, 1+\epsilon; \frac{\alpha_1 \alpha_3 + \alpha_2 \alpha_4 - \alpha_2 \alpha_3}{\alpha_2 (\alpha_4 - \alpha_3)} \right) \right. \\ & \left. -((\alpha_1-\alpha_2) (\alpha_4 - \alpha_3))^{-\epsilon} \; _2F_1 \left( \epsilon, \epsilon, 1+\epsilon; -\frac{\alpha_1 \alpha_3 + \alpha_2 \alpha_4 - \alpha_2 \alpha_3}{(\alpha_1-\alpha_2) (\alpha_4 - \alpha_3)} \right) \right]\; ,
\end{split}
\end{equation}
where
\begin{equation}
    r_{\Gamma} = \frac{\Gamma(1-\epsilon) \Gamma^2(1+\epsilon)}{\Gamma(1+2 \epsilon)} \; .
\end{equation}
It is clear that with appropriate exchanges of the invariants
$s,s_1,s_2,t_1,t_2$, they can be obtained from each other. In particular, as
pointed out in~\cite{Bern_1994}, starting from one of these integrals, the
others can be obtained by cyclic permutations of the $\alpha_i$'s. 

By expanding these results, in multi-Regge kinematics, we have
\begin{equation}
   \hat{I}_{4}^{(1)} (s_1,s_2,s) \simeq \frac{ \Gamma(1-\epsilon) \Gamma^2(1+\epsilon)}{\Gamma(1+2 \epsilon)} \frac{2}{\epsilon^2} \left( \frac{(-s_1)(-s_2)}{(-s)} \right)^{\epsilon}  \left \{ 1 + \sum_{n=1}^{\infty} \epsilon^{2n} \;  2 \left( 1 - \frac{1}{2^{2 n -1}} \right) \zeta (2 n) \right \} \; ,
\end{equation}
\begin{gather}
    \hat{I}_{4}^{(2)} (s_2,t_2,t_1) =  \frac{\Gamma(1-\epsilon) \Gamma^2(1+\epsilon)}{\Gamma(1+2 \epsilon)} \frac{2}{\epsilon^2} \left(-t_1\right)^\epsilon \left[ \left(\frac{t_2}{t_1}\right)^\epsilon \bigg ( e^{-i \pi \epsilon} \Gamma(1-\epsilon) \Gamma(1+\epsilon) \right. \nonumber \\
    \left. \left. + 1  -\epsilon \ln \left( \frac{ t_1-t_2 }{s_2} \right) -\sum_{n=2}^{\infty}(-\epsilon)^n \zeta(n) \right ) - 1 + \epsilon \ln \left(1- \frac{t_1}{t_2} \right) + \sum_{n=2}^{\infty}(-\epsilon)^n {\rm{Li}}_n\left( \frac{t_1}{ t_2} \right) \right ] \; .
\end{gather}
\begin{equation}
\hat{I}_{4}^{(3)} (s,s_1,t_2) \simeq \frac{ \Gamma(1-\epsilon) \Gamma^2(1+\epsilon)}{\Gamma(1+2 \epsilon)} \frac{2}{\epsilon^2} (-t_2)^{\epsilon} \left \{ 1 + \epsilon \ln \left( \frac{-s}{-s_1} \right) - \sum_{n=2}^{\infty} (-\epsilon)^{n} \zeta(n) \right \} \; ,
\end{equation}
\begin{equation}
\hat{I}_{4}^{(4)} (s,s_2,t_1) \simeq \frac{ \Gamma(1-\epsilon) \Gamma^2(1+\epsilon)}{\Gamma(1+2 \epsilon)} \frac{2}{\epsilon^2} (-t_1)^{\epsilon} \left \{ 1 + \epsilon \ln \left( \frac{-s}{-s_2} \right) - \sum_{n=2}^{\infty} (-\epsilon)^{n} \zeta(n) \right \} \; ,
\end{equation}
\begin{gather}
    \hat{I}_{4}^{(5)} (s_1,t_1,t_2) =  \frac{\Gamma(1-\epsilon) \Gamma^2(1+\epsilon)}{\Gamma(1+2 \epsilon)} \frac{2}{\epsilon^2} \left(-t_2\right)^\epsilon \left[ \left(\frac{t_1}{t_2}\right)^\epsilon \bigg ( e^{-i \pi \epsilon} \Gamma(1-\epsilon) \Gamma(1+\epsilon) \right. \nonumber \\
    \left. \left. + 1  -\epsilon \ln \left( \frac{ t_2-t_1 }{s_1} \right) -\sum_{n=2}^{\infty}(-\epsilon)^n \zeta(n) \right ) - 1 + \epsilon \ln \left(1- \frac{t_2}{t_1} \right) + \sum_{n=2}^{\infty}(-\epsilon)^n {\rm{Li}}_n\left( \frac{t_2}{t_1} \right) \right ] \; .
\end{gather} 
Again, truncating the summation at $n=4$ gives the desired result. The integral $\hat{I}_4^{(2)}$, up to trivial factors, coincides with~(\ref{I4B_BDK}), as can be easily verified. Since we have computed the integral $\mathcal{I}_{4B}$ through two independent methods, the integral $\hat{I}_4^{(2)}$ has been cross-checked. We can use the technique explained in Appendix~\ref{Appendix C} to obtain the relation (\ref{I4B-I4Bcal}) and then, by using the knowledge of $\mathcal{I}_{4B}$ integral from the alternative method explained in Section~\ref{subsec:I42}, obtain an alternative expression for $\hat{I}_4^{(2)}$. This automatically verifies $\hat{I}_4^{(5)}$, that is obtained from $\hat{I}_4^{(2)}$ by the substitutions $s_2 \rightarrow s_1$ and $t_2 \leftrightarrow t_1$. The technique explained in Appendix~\ref{Appendix C} can be also applied to $\hat{I}_4^{(1)}$, $\hat{I}_4^{(3)}$, $\hat{I}_4^{(4)}$ in order to verify also the results. These latter cases are really much simpler with respect to the ones explicitly computed in Appendix~\ref{Appendix C}. We can also verify these integrals using direct the Feynman technique, as explained in~\ref{Appendix B} for $I_{4B}$. 

\subsubsection{Del Duca, Duhr, Glover, Smirnov result for the $\hat{I}_5^{D = 6 + 2 \epsilon}$ part}
\label{subsubsec:DDGS}
As it can be seen from eq.~(\ref{IterativeRelI5}), beyond the constant order in the $\epsilon$-expansion, the pentagon in dimension $D=4+2 \epsilon$ takes a contribution from the pentagon integral in dimension $D=6 + 2 \epsilon$. The latter pentagon integral is finite and hence does not contribute to the divergent and finite parts in~(\ref{IterativeRelI5}). It starts to contribute at the order $\epsilon$, and hence, if one wants to obtain the pentagon in $D=4+2 \epsilon$ up to $\epsilon^2$ accuracy, the first two non-trivial orders of the pentagon in $D=6 + 2 \epsilon$ must be computed. There are different results in literature for the 6-dimensional pentagon integral~\cite{DelDuca:2009ac,Kniehl:2010aj,Kozlov:2015kol,Syrrakos:2020kba}; among these, the one obtained by Del Duca, Duhr, Glover, Smirnov (DDGS) is the most suitable for our aims, as it is calculated in multi-Regge kinematics and its $\epsilon$-expansion up to the desired order is given. In~\cite{DelDuca:2009ac} the integral is computed by two independent methods: 1) negative dimension approach and 2) Mellin-Barnes technique. The final result can be expressed in terms of transcendental double sums that, following Ref.~\cite{DelDuca:2009ac}, we denote by $\mathcal{M}$-functions; their definition is given in Appendix~\ref{AppendixA}. Adapting their notation to ours, we define
\begin{equation}
    x_1 \equiv \frac{s t_1}{s_1 s_2} = \frac{t_1}{\vec{p}^{ \; 2}}  \; , \hspace{0.5 cm} x_2 = \frac{s t_2}{s_1 s_2} = \frac{t_2}{\vec{p}^{ \; 2}} \; , \hspace{0.5 cm} - \vec{p}^{\; 2} = \frac{(-s_1)(-s_2)}{(-s)} \; , \hspace{0.5 cm} y_1 \equiv \frac{1}{x_2} \; , \hspace{0.5 cm} y_2 = \frac{x_1}{x_2} \; .
\end{equation}
In Ref.~\cite{DelDuca:2009ac} the three regions contributing to the pentagonal are identified as
\begin{itemize}
    \item \textbf{Region I} : $\sqrt{x_1} + \sqrt{x_2} < 1$ 
    \item \textbf{Region II (a)} : $-\sqrt{x_1} + \sqrt{x_2} > 1$ 
    \item \textbf{Region II (b)} : $\sqrt{x_1} - \sqrt{x_2} > 1$ 
\end{itemize}
The solution in the \textbf{Region II (a)} is
\begin{equation}
    \hat{I}_{5}^{(D=6+2 \epsilon)} = -  \frac{ \Gamma(1-\epsilon) \Gamma^2(1+\epsilon)}{\Gamma(1+2 \epsilon)} \sqrt{- \frac{s_1 s_2 t_1}{s t_2}} \left( \frac{(-s_1)(-s_2)}{(-s)} \right)^{\epsilon} \mathcal{I}_{DDGS}^{II(a)} (\vec{p}^{\; 2}, t_1, t_2) \; ,
    \label{DDGSsol}
\end{equation}
where
\begin{equation}
   \mathcal{I}_{DDGS}^{II(a)} (\vec{p}^{\; 2}, t_1, t_2) = i_0^{II(a)} (y_1, y_2) - \epsilon \; i_1^{II(a)} (y_1, y_2) + \mathcal{O} (\epsilon^2)\;,
\end{equation}
with 
\begin{align}
   & i^{(IIa)}_0 (y_1,y_2)=
(-8 \ln y_1-4 \ln y_2) \cM\big(0,0,(1,1);-y_1, y_2\big)-4 \ln y_2 \cM\big((1,1),0,0;-y_1, y_2\big) \nonumber \\
+&\,18 \cM\big(0,0,(1,2);-y_1, y_2\big)+18 \cM\big(0,0,(2,1);-y_1, y_2\big)-24 \cM\big(0,0,(1,1,1);-y_1, y_2\big) \nonumber \\
+&\,8 \cM\big(0,1,(1,1);-y_1, y_2\big)+16 \cM\big(1,0,(1,1);-y_1, y_2\big)-8 \cM\big((1,1),0,1;-y_1, y_2\big) \nonumber \\
+&\,8 \cM\big((1,1),1,0;-y_1, y_2\big)-\cM\big(0,0,0;-y_1, y_2\big) \Big(\frac{\pi ^2 \ln y_1}{3}+\frac{\ln^2 y_1 \ln y_2}{2}+\frac{\pi ^2 \ln y_2}{2}-2 \zeta(3)\Big) \nonumber \\
-&\,\cM\big(0,0,1;-y_1, y_2\big) \Big(2 \ln y_1 \ln y_2+\ln^2 y_1+\frac{5 \pi ^2}{3}\Big)+(6 \ln y_1+3 \ln y_2) \cM\big(0,0,2;-y_1, y_2\big) \nonumber \\
+&\,\Big(2 \ln y_1 \ln y_2+\frac{2 \pi ^2}{3}\Big) \cM\big(1,0,0;-y_1, y_2\big)+(4 \ln y_1+4 \ln y_2) \cM\big(1,0,1;-y_1, y_2\big) \nonumber \\
+&\,4 \ln y_1 \cM\big(0,1,1;-y_1, y_2\big)-4 \ln y_1 \cM\big(1,1,0;-y_1, y_2\big)+\Big(\ln^2 y_1+\pi ^2\Big) \cM\big(0,1,0;-y_1, y_2\big) \nonumber \\
+&\,\ln y_2 \cM\big(2,0,0;-y_1, y_2\big)-12 \cM\big(0,0,3;-y_1, y_2\big)-6 \cM\big(0,1,2;-y_1, y_2\big) \nonumber \\
-&\,12 \cM\big(1,0,2;-y_1, y_2\big)-8 \cM\big(1,1,1;-y_1, y_2\big)+2 \cM\big(2,0,1;-y_1, y_2\big) \nonumber \\
-&\,2 \cM\big(2,1,0;-y_1, y_2\big),
\end{align}
\begin{align}
& i^{(IIa)}_1 (y_1,y_2)=
\cM\big(0,0,(1,1);-y_1, y_2\big) \Big(4 \ln y_1 \ln y_2-4 \ln^2 y_1+2 \ln^2 y_2+4 \pi ^2\Big) \nonumber \\
+&\,(2 \ln^2 y_2-4 \ln y_1 \ln y_2) \cM\big((1,1),0,0;-y_1, y_2\big)+(8 \ln y_1-12 \ln y_2) \cM\big(1,0,(1,1);-y_1, y_2\big) \nonumber \\
+&\,(4 \ln y_2-8 \ln y_1) \cM\big((1,1),0,1;-y_1, y_2\big)+(8 \ln y_1-4 \ln y_2) \cM\big((1,1),1,0;-y_1, y_2\big) \nonumber \\
-&\,15 \ln y_2 \cM\big(0,0,(1,2);-y_1, y_2\big)-15 \ln y_2 \cM\big(0,0,(2,1);-y_1, y_2\big) \nonumber \\
+&\,20 \ln y_2 \cM\big(0,0,(1,1,1);-y_1, y_2\big)-4 \ln y_2 \cM\big(0,1,(1,1);-y_1, y_2\big) \nonumber \\
-&\,\ln y_2 \cM\big((1,2),0,0;-y_1, y_2\big)-\ln y_2 \cM\big((2,1),0,0;-y_1, y_2\big) \nonumber \\ +&\,4 \ln y_2 \cM\big((1,1,1),0,0;-y_1, y_2\big)+32 \cM\big(0,0,(1,3);-y_1, y_2\big)+36 \cM\big(0,0,(2,2);-y_1, y_2\big) \nonumber \\
+&\,32 \cM\big(0,0,(3,1);-y_1, y_2\big)-48 \cM\big(0,0,(1,1,2);-y_1, y_2\big)-48 \cM\big(0,0,(1,2,1);-y_1, y_2\big) \nonumber \\
-&\,48 \cM\big(0,0,(2,1,1);-y_1, y_2\big)+64 \cM\big(0,0,(1,1,1,1);-y_1, y_2\big)+12 \cM\big(0,1,(1,2);-y_1, y_2\big) \nonumber \\
+ & 12 \cM\big(0,1,(2,1);-y_1, y_2\big)-16 \cM\big(0,1,(1,1,1);-y_1, y_2\big)+18 \cM\big(1,0,(1,2);-y_1, y_2\big) \nonumber \\
+&\,18 \cM\big(1,0,(2,1);-y_1, y_2\big)-24 \cM\big(1,0,(1,1,1);-y_1, y_2\big)+8 \cM\big(1,1,(1,1);-y_1, y_2\big) \nonumber \\
-&\,2 \cM\big((1,2),0,1;-y_1, y_2\big)+2 \cM\big((1,2),1,0;-y_1, y_2\big)-2 \cM\big((2,1),0,1;-y_1, y_2\big) \nonumber \\
+&\,2 \cM\big((2,1),1,0;-y_1, y_2\big)+8 \cM\big((1,1,1),0,1;-y_1, y_2\big)-8 \cM\big((1,1,1),1,0;-y_1, y_2\big) \nonumber \\
+&\,\cM\big(0,0,1;-y_1, y_2\big) \Big(\ln y_1 \ln^2 y_2-\frac{\pi ^2 \ln y_1}{3}-\frac{\ln^2 y_1 \ln y_2}{2}-\frac{2 \ln^3 y_1}{3}+\frac{3 \pi ^2 \ln y_2}{2}-6 \zeta(3)\Big) \nonumber \\
+&\,\cM\big(0,1,0;-y_1, y_2\big) \Big(\frac{\pi ^2 \ln y_1}{3}-\frac{\ln^2 y_1 \ln y_2}{2}+\frac{2 \ln^3 y_1}{3}-\frac{\pi ^2 \ln y_2}{2}+2 \zeta(3)\Big) \nonumber \\
+&\,\cM\big(1,0,0;-y_1, y_2\big) \Big(-\ln y_1 \,\ln^2 y_2+\frac{\pi ^2 \ln y_1}{3}+\frac{3 \ln^2 y_1 \ln y_2}{2}-\frac{\pi ^2 \ln y_2}{2}+2 \zeta(3)\Big) \nonumber \\
+&\,\cM\big(0,0,2;-y_1, y_2\big) \Big(-3 \ln y_1 \ln y_2+3 \ln^2 y_1-\frac{3 \ln^2 y_2}{2}-3 \pi ^2\Big) \nonumber \\
+&\,\cM\big(0,1,1;-y_1, y_2\big) \Big(-2 \ln y_1 \ln y_2+2 \ln^2 y_1-\frac{4 \pi ^2}{3}\Big) \nonumber \\
+&\,\cM\big(1,1,0;-y_1, y_2\big) \Big(2 \ln y_1 \ln y_2-3 \ln^2 y_1+\frac{\pi ^2}{3}\Big)+(\ln y_2-2 \ln y_1) \cM\big(2,1,0;-y_1, y_2\big) \nonumber \\
+&\,\cM\big(2,0,0;-y_1, y_2\big) \Big(\ln y_1 \ln y_2-\frac{\ln^2 y_2}{2}\Big)+(9 \ln y_2-6 \ln y_1) \cM\big(1,0,2;-y_1, y_2\big) \nonumber \\
+&\,(4 \ln y_2-4 \ln y_1) \cM\big(1,1,1;-y_1, y_2\big)+(2 \ln y_1-\ln y_2) \cM\big(2,0,1;-y_1, y_2\big) \nonumber \\
+&\,\cM\big(0,0,0;-y_1, y_2\big) \Big(\frac{\ln^2 y_1 \ln^2 y_2}{4}-\frac{\pi ^2 \ln^2 y_1}{6}-\frac{\ln^3 y_1 \ln y_2}{3}-2 \ln y_2 \zeta(3)+\frac{\pi ^2 \ln^2 y_2}{4}+\frac{2 \pi ^4}{15}\Big) \nonumber \\
+&\,\Big(3 \ln^2 y_1-2 \ln^2 y_2-\pi ^2\Big) \cM\big(1,0,1;-y_1, y_2\big)+10 \ln y_2 \cM\big(0,0,3;-y_1, y_2\big) \nonumber \\
+&\,3 \ln y_2 \cM\big(0,1,2;-y_1, y_2\big)-20 \cM\big(0,0,4;-y_1, y_2\big)-8 \cM\big(0,1,3;-y_1, y_2\big) \nonumber \\
-&\,12 \cM\big(1,0,3;-y_1, y_2\big)-6 \cM\big(1,1,2;-y_1, y_2\big).
\end{align}
The solution in the \textbf{Region II (b)} is obtained by the replacement
\begin{equation}
   \mathcal{I}_{DDGS}^{II(a)} (\vec{p}^{\; 2}, t_1, t_2) \rightarrow \mathcal{I}_{DDGS}^{II(b)} (\vec{p}^{\; 2}, t_1, t_2) = \frac{t_2}{t_1} \mathcal{I}_{DDGS}^{II(a)} (\vec{p}^{\; 2}, t_2, t_1) 
\end{equation}
in eq.~(\ref{DDGSsol}). \\
The solution in the \textbf{Region I} can be written as  
\begin{equation}
    \hat{I}_{5}^{(6+2 \epsilon)} = -  \frac{ \Gamma(1-\epsilon) \Gamma^2(1+\epsilon)}{\Gamma(1+2 \epsilon)} \sqrt{- \frac{s t_1 t_2}{s_1 s_2}} \left( \frac{(-s_1)(-s_2)}{(-s)} \right)^{\epsilon} \mathcal{I}_{DDGS}^{I} (\vec{p}^{\; 2}, t_1, t_2) \; ,
\end{equation}
where $\mathcal{I}_{DDGS}^{I} (\vec{p}^{\; 2}, t_1, t_2)$ can be obtained from $\mathcal{I}_{DDGS}^{II(a)} (\vec{p}^{\; 2}, t_1, t_2)$ by analytical continuation, according to the prescription $y_1 \rightarrow 1/y_1$.

\section{Conclusions and outlook}
\label{Conclusions}

In this paper we have calculated at the NLO the Reggeon-Reggeon-gluon (also called ``Lipatov") effective vertex in QCD with accuracy up to the order $\epsilon^2$, with $\epsilon=(D-4)/2$ and $D$ the space-time dimension. The NLO Lipatov effective vertex can be expressed in terms of a few integrals (triangle, boxes, pentagon), which we obtained at the required accuracy in a two-fold way: 1) taking their expressions, known at arbitrary $\epsilon$, from the literature~\cite{Bern_1994,DelDuca:2009ac} and expanding them to the required order; 2) calculating them from scratch\footnote{With the exception of the part of the pentagon integral in $D = 4+2 \epsilon$ which depends on the same integral in $D = 6+2 \epsilon$. Nevertheless, in Ref.~\cite{DelDuca:2009ac} this contribution was computed through two independent methods. In the case of $\mathcal{I}_3$ we have used an independent method valid just up to the order $\epsilon^2$.}, by an independent method. The purpose of the latter calculation is not only cross-checking, but also providing with an alternative, though equivalent, expression for the integrals, which could turn out to be more convenient for the uses of the NLO Lipatov effective vertex. For instance, the result up to order $\epsilon$ of the integral $\mathcal{I}_3$, calculated in Section~\ref{subsec:I32}, is very compact compared to the result for the same integral computed in~\ref{subsec:I31}. It contains the structure $I_{\vec{q}_1^{\;2},\vec{q}_2^{\;2},\vec{p}^{\;2}}$ which has been extensively used in the BFKL literature (see \textit{e.g.}~\cite{Ioffe:2010zz}).  \\

The integrals $\mathcal{I}_{4B}$, $\mathcal{I}_{4A}$ and $\hat{I}_{4}^{(i)}$ have been written as expansions to all orders in $\epsilon$, in terms of polylogarithms. The integral $\mathcal{I}_3$ has been expanded up to the order $\epsilon^2$, but can be expanded to higher orders, if needed. The residual and most complicated term is the one related to the pentagonal integral in dimension $6-2\epsilon$. Its expression to the order $\epsilon^2$ in MRK is given in Section~\ref{subsubsec:DDGS}. The knowledge of the one-loop Lipatov vertex in QCD at any successive order in the $\epsilon$-expansion is completely reduced to the computation of this integral with higher $\epsilon$-accuracy. \\

There are a number of reasons motivating the need of the NLO Lipatov effective vertex with higher $\epsilon$ accuracy: first, it is the building block of the next-to-NLO contribution to the BFKL kernel from the production of one gluon in the collision of two Reggeons; second, it enters the expression of the impact factors for the Reggeon-gluon transition, which appear in the derivation of the bootstrap conditions for inelastic amplitudes; these discontinuities are needed in the derivation of the BFKL equation in the NNLA; third, the discontinuities of multiple gluon production amplitudes in the MRK can be used for a simple demonstration of violation of the ABDK-BDS (Anastasiou-Bern-Dixon-Kosower — Bern-Dixon-Smirnov) ansatz for amplitudes with maximal helicity violation in Yang-Mills theories with maximal super-symmetry ($\mathcal{N}=4$ SYM) in the planar limit and for the calculations of the remainder functions to this ansatz. 

\newpage

\appendix

\section{Polylogarithms, Hypergeometric functions and Nested harmonic sums}
\label{AppendixA}
In this Appendix we give some usual definitions and useful relations. \vspace{0.3 cm} 

\textit{\bf Polylogarithms}~\cite{kolbig:1986ngp}\\

We define polylogarithms as
\begin{equation}
{\rm{Li}}_{a+1}(z) = \frac{(-1)^a \; z}{a!} \int_0^1 dt \frac{\ln^a (t)}{1-t z} 
\end{equation}
and Nielsen generalized polylogarithms as 
\begin{equation}
    {\rm{S}}_{a,b}(z) = \frac{(-1)^{a+b-1}}{(a-1)!b!} \int_0^1 d t \frac{\ln^{a-1} t \ln^{b} (1 - z t)}{t} \; , \hspace{0.5 cm} {\rm{S}}_{a,1}(z) = {\rm{Li}}_{a+1}(z) \; \, ,
\end{equation}
where $a$ and $b$ are integers. During calculations, the following inversion and reflection formulas for polylogarithms are useful~\cite{kolbig:1986ngp},
\begin{equation*}
    {\rm{Li}}_n (z) + (-1)^n {\rm{Li}}_n \left( \frac{1}{z} \right) = -\frac{1}{n!} \ln^n (-z) - \sum_{j=0}^{n-2} \frac{1}{j!} (1+(-1)^{n-j})(1-2^{1-n+j}) \zeta (n-j) \ln^{j} (-z) \; ,
\end{equation*}
\begin{equation*}
    {\rm{S}}_{n,p} (z) = \sum_{j=0}^{n-1} \frac{\ln^j z}{j!} \left \{ s_{n-j, p} - \sum_{k=0}^{p-1} \frac{(-1)^k \ln^k (1-z)}{k!} {\rm{S}}_{p-k, n-j} (1-z) \right \} + \frac{(-1)^p}{n!p!} \ln^n z \ln^p (1-z) \, ,
\end{equation*}
where 
\begin{equation*}
   s_{n,p} = S_{n,p} (1) \; . 
\end{equation*}
In particular, from these relations, we obtain
\begin{equation*}
    {\rm{Li}}_2 (z) = - {\rm{Li}}_2 \left(\frac{1}{z} \right) - \zeta(2) - \frac{1}{2} \ln^2 (-z) \; , \hspace{0.5 cm} {\rm{Li}}_3 (z) = {\rm{Li}}_3 \left(\frac{1}{z} \right) - \zeta(2) \ln (-z) - \frac{1}{6} \ln^3 (-z) \; ,
\end{equation*}
\begin{equation*}
    {\rm{Li}}_4 (z) = - {\rm{Li}}_4 \left(\frac{1}{z} \right) - \frac{7}{4} \zeta (4) - \frac{1}{2} \zeta(2) \ln^2 (-z) - \frac{1}{4!} \ln^4 (-z) \; , 
\end{equation*}
\begin{equation*}
    {\rm{Li}}_2 (z) = \zeta(2) - {\rm{Li}}_2 (1-z) - \ln z \ln(1-z)  \; , 
\end{equation*}
\begin{equation*}
    {\rm{Li}}_3 (z) = - {\rm{Li}}_3 (1-z) - {\rm{Li}}_3 \left(1-\frac{1}{z} \right) + \zeta (3) + \frac{1}{6} \ln^3 z + \zeta (2) \ln z - \frac{1}{2} \ln^2 z \ln (1-z) \, ,
\end{equation*}
\begin{equation*}
   \text{Li}_4 \left(z \right) = \zeta(4) - \text{S}_{1,3} (1-z) + \ln z ( \zeta (3) - \text{S}_{1,2} (1-z)) + \frac{\ln^2 z}{2} (\zeta(2) - \text{Li}_2 (1-z)) - \frac{1}{6} \ln^3 z \ln (1-z) \; .
\end{equation*}
It is also useful to know that,
\begin{equation}
    \frac{d}{d y} {\rm{Li}}_2 \left( y \right) = - \frac{\ln (1-y)}{y} \; , \hspace{0.5 cm} \frac{d}{d y} {\rm{Li}}_3 \left( y \right) = \frac{{\rm{Li}}_2 \left( y \right)}{y} \; , \hspace{0.5 cm} \frac{d}{d y} {\rm{S}}_{1,2} \left( y \right) = \frac{\ln^2 (1-y)}{2 y} \; ,
\end{equation}
\begin{equation}
    \Im \{ {\rm{Li}}_2 (y + i \varepsilon) \} = \pi \theta (y-1) \ln (y) \; , \hspace{0.5 cm} \Im \{ {\rm{Li}}_3 (y + i \varepsilon) \} = \pi \theta (y-1) \frac{\ln^2 (y)}{2} \;,
\end{equation}
\begin{equation}
    \Im \{ {\rm{S}}_{1,2} (y + i \varepsilon) \} = \pi \theta (y-1) \left[ \zeta (2) - {\rm{Li}}_2 \left( \frac{1}{y} \right) - \frac{\ln^2 (y)}{2} \right]  \;.
\end{equation}

\phantom{.}\\
\textit{\bf Hypergeometric function $_2 F_1$} \\

We can represent the hypergeometric function $_2 F_1$ as
\begin{equation}
    _2 F_1 (a,b,c;z) = \frac{1}{B(b,c-b)} \int_0^1 dx \; x^{b-1} (1-x)^{c-b-1} (1 - z x)^{-a} \; ,
\label{IntReprHyperGau}
\end{equation}
for $\Re \{ c \} > \Re \{ b \} >0 $. The definition is valid in the entire complex $z$-plane with a cut along the real axis from one to infinity.
Using the integral representation (\ref{IntReprHyperGau}) and performing the transformation
\begin{equation}
    x = - \frac{y}{z \left[ 1 - \left( 1 + \frac{1}{z} \right) y \right]} \; ,
\end{equation}
one can prove the following identity
\begin{equation}
_2F_1 (a, b, 1 + b; 1+ z ) = (-z)^{-b} \; _2 F_1 \left( 1+b-a, b, 1+b ; 1 + z^{-1} \right) \; .
\label{HyperMoreGenProp}
\end{equation}
Choosing $a=b=\epsilon$, we get
\begin{equation}
_2 F_1 (1, \epsilon, 1+\epsilon ; 1 + z) =\; (-z)^{-\epsilon} \; _2F_1 (\epsilon, \epsilon, 1+\epsilon ; 1+z^{-1}) \; .
\label{HyperProp1}
\end{equation}
One can also prove the following expansions~\cite{Kalmykov:2006hyp}:
\begin{equation}
_2 F_1 (1, -\epsilon, 1-\epsilon ; z) = 1 - \sum_{i=1}^{\infty} \epsilon^i {\rm{Li}}_i(z) \; ,
\label{HyperExp2}
\end{equation}
\begin{equation*}
    _2 F_1(2 - 2 \epsilon, 1-\epsilon, 2-\epsilon;z) = \frac{1}{1-z} \left \{ 1 + \left[ 1 + \left( \frac{1+z}{z} \right) \ln(1-z) \right] \epsilon \right.
\end{equation*}
\begin{equation*}
    + \left[ 2 + \left( \frac{1+z}{z} \right)  \ln (1-z) \ln ((1-z)e) - \left( \frac{1-z}{z} \right) {\rm{Li}}_2(z)  \right] \epsilon^2
\end{equation*}
\begin{equation*}
  \left. + \left[ 4 + \ln (1-z) \left( \frac{2(1+z)}{z} + \frac{2(1-z)}{z} \zeta(2) + \frac{\ln (1-z)}{3z} \right.  \right. \right.  
\end{equation*}
\begin{equation*}
   \left. \times \bigg( 3(1+z) + 2(1+z) \ln (1-z) - 3 (1-z) \ln z \right) \bigg) - \frac{1-z}{z} {\rm{Li}}_2 (z) \left( 1+ 2 \ln (1-z) \right)  
\end{equation*}
\begin{equation}
  \left. \left. - \frac{2 (1-z)}{z} \left( {\rm{Li}}_3 (1-z) + \frac{{\rm{Li}}_3 (z)}{2} - \zeta (3) \right) \right] \epsilon^3 \right \} + \mathcal{O}(\epsilon^4) \; .
\label{HyperExp1}
\end{equation}

\phantom{.}\\
\textit{\bf Nested harmonic sums and $\mathcal{M}$ functions}~\cite{DelDuca:2009ac} \vspace{0.2 cm} \\
The nested harmonic sums are defined recursively by
\begin{equation}
    S_i (n) = \sum_{k=1}^n \frac{1}{k^i} \; , \hspace{1 cm} S_{i \vec{j}} (n) = \sum_{k=1}^{n} \frac{S_{\vec{j}} (k)}{k^i} \; ,
\end{equation}
while the $\mathcal{M}$-functions are defined by the double series
\begin{equation}
    \mathcal{M} (\vec{i}, \vec{j}, \vec{k}; x_1, x_2) = \sum_{n_1=0}^{\infty} \sum_{n_2 =0}^{\infty} \binom{n_1+n_2}{n_1}^2 S_{\vec{i}} (n_1) S_{\vec{j}} (n_2) S_{\vec{k}} (n_1+n_2) x_1^{n_1} x_2^{n_2} \; .
\end{equation}
\section{Some useful integrals}
\label{Appendix B}

\textbf{The integral $I_{a,b,c}$} 
\vspace{0.2 cm}

The integral
\begin{equation}
    I_{a,b,c} = \int_0^1 dx \ \frac{1}{a x + b (1-x) - c x (1-x)} \ln \left( \frac{a x + b(1-x) }{c x(1-x)} \right) 
\tag{\ref{FadinGorba}}
\end{equation}
is invariant with respect to any permutation of its arguments, as it can be seen from the representation
\begin{equation}
    I_{a,b,c} = \int_0^1 d x_1 \int_0^1 d x_2 \int_0^1 d x_3 \ \frac{\delta(1-x_1-x_2-x_3)}{(a x_1 + b x_2 + c x_3)(x_1 x_2 + x_1 x_3 + x_2 x_3)} \;.
    \label{Isimm}
\end{equation}
To prove that (\ref{FadinGorba}) and (\ref{Isimm}) are equivalent, we first integrate over $x_3$ and then perform the change of variables
\begin{equation*}
    x = \frac{x_2}{x_1+x_2} \; , \hspace{0.5 cm} z= \frac{x_1 x_2}{(1-x_1)x_1 + (1-x_2)x_2 -x_1 x_2} \; ,
\end{equation*}
\begin{equation*}
    x_1 = \frac{(1-x)z}{z+x(1-x)(1-z)} \; , \hspace{0.5 cm} x_2 = \frac{x z}{z+x(1-x)(1-z)} \; , \hspace{0.5 cm}  x_3 = \frac{x(1-x)(1- z)}{z+x(1-x)(1-z)} ,
\end{equation*}
with Jacobian
\begin{equation}
    J =  \frac{xz(1-x)}{[z+x(1-x)(1-z)]^{3}} \; .
\end{equation}
We obtain
\begin{equation*}
    I_{a,b,c} = \int_0^1 d x \int_0^{\infty} d z \ \frac{\Theta \left( \frac{x(1-x)(1-z)}{z+x(1-x)(1-z)} \right)}{a z(1-x)  + b x z  + c x(1-x)(1-z) } 
\end{equation*}
\begin{equation}
    = \int_0^1 dx \  \frac{1}{a (1-x) + b x - c x(1-x)} \ln \left( \frac{b  x + a (1-x)}{c x (1-x) }\right) \; ,
\end{equation}
which is equal to (\ref{FadinGorba}) after the trivial change of variables $x \leftrightarrow 1-x$. \\
Another useful representation of immediate proof is 
\begin{equation}
    I_{a,b,c} = \int_0^1 dx \int_1^{\infty} dt \ \frac{1}{t\ [a x(1-x)(t-1) + b (1-x) + c x ]} \vspace{0.2 cm} \; .
\end{equation}

In the case when $a=\vec q_1^{\;2}$, $b=\vec q_2^{\;2}$ and $c=(\vec q_1-\vec q_2)^{2}\equiv \vec p^{\;2}$, the explicit solution of the integral is~\cite{Fadin:2000kx}
\begin{equation*}
    I_{\vec{q}_1^{\;2},\vec{q}_2^{\;2},\vec{p}^{\;2}} = \int_0^1 dx \frac{1}{ \vec{q}_1^{\; 2} x + \vec{q}_2^{\; 2} (1-x) - \vec{p}^{\; 2} x(1-x)} \ln \left( \frac{\vec{q}_1^{\; 2} x + \vec{q}_2^{\; 2} (1-x)}{\vec{p}^{\; 2}x(1-x)} \right) 
\end{equation*}
\begin{equation}
    = - \frac{2}{|\vec{q}_1| |\vec{q}_2| \sin{\phi}} \left[ \ln \rho \arctan \left( \frac{\rho \sin \phi}{1-\rho \cos \phi} \right) + \Im \left( - {\rm{Li}}_2 (\rho e^{i \phi}) \right) \right] \; , 
    \label{Isol}
\end{equation}
where $\phi$ is the angle between $\vec{q}_1$, $\vec{q}_2$ and $\rho = {\rm{min}} \left( \frac{|\vec{q}_1|}{|\vec{q}_2|}, \frac{|\vec{q}_2|}{|\vec{q}_1|}\right)$. \vspace{1.5 cm} \\
\textbf{The box integral with one external mass} 
\vspace{0.3 cm} \\
Here, we derive the result for $I_{4B}$ in Eq.~(\ref{C}), \textit{i.e.} a box integral with massless propagators and one external mass, using direct Feynman technique. The integral is
\begin{equation}
  I_{4B} = \frac{1}{i} \int d^{D}k \frac{1}{(k^{2} + i \varepsilon) [(k+q_1)^{2}+i\varepsilon]
    [(k+q_2)^{2}+i\varepsilon] [(k-p_B)^2 + i \varepsilon]} \; .
  \tag{\ref{I4B}}
\end{equation}
Defining 
\begin{equation}
    d_1 = k^{2} + i \varepsilon \; , \hspace{0.5 cm} d_2 = (k+q_1)^{2}+i\varepsilon \; , \hspace{0.5 cm} d_3 = (k+q_2)^{2}+i\varepsilon \; , \hspace{0.5 cm} d_4 = (k-p_B)^2 + i \varepsilon \; ,
\end{equation}
we can write
\begin{equation}
\frac{1}{d_1 d_4}=\int_0^1 d x \frac{1}{\left((1-x) d_1+x d_4\right)^2} \; , \hspace{0.5 cm} \frac{1}{d_2 d_3}=\int_0^1 d y \frac{1}{\left((1-y) d_2+y d_3\right)^2} \; ,
\end{equation}
and hence
\begin{equation}
    \frac{1}{d_1 d_2 d_3 d_4}=\Gamma(4) \int_0^1 dy \int_0^1 dx \int_0^1 dz \frac{ z (1-z) }{\left[(1-z)\left((1-x) d_1+x d_4\right)+z\left((1-y) d_2+y d_3\right)\right]^4} \; .
\end{equation}
After the integration in the $d^D k$, we obtain 
\begin{equation*}
    I_{4 B}=\pi^{D / 2} \Gamma(4-D / 2)  
\end{equation*}
\begin{equation}
 \times \int_0^1  d z \int_0^1 d x \int_0^1 d y \frac{z(1-z)}{ \left[z(1-z)\left(-s_2 x(1-y)+(b y+a(1-y))(1-x) \right)-i 0 \right]^{4-D/2} } \; ,
\end{equation}
where $a=-t_1, \; b= -t_2$. Performing the trivial integrations over $z$ and $x$, we have
\begin{equation*}
I_{4 B}=\pi^{D / 2} \Gamma(1-\epsilon) \frac{\Gamma^2(\epsilon)}{\Gamma(2 \epsilon)} 
\end{equation*}
\begin{equation}
   \times \int_0^1 \frac{d y}{s_2(1-y)+b y+a(1-y)}\left[\left(-s_2(1-y)-i 0\right)^{\epsilon-1}-(b y+a(1-y))^{\epsilon-1}\right] .
\end{equation}
The first term in the square bracket gives
\begin{equation*}
\left(-s_2-i0\right)^{\epsilon-1} \int_0^1 dy \frac{(1-y)^{\epsilon-1}}{\left(s_2+a\right)(1-y)+b y} =\left(-s_2-i0 \right)^{\epsilon-1} \frac{1}{b} \int_0^1 dx \frac{x^{\epsilon-1}}{1-x\left(1- \left(s_2+a\right) / b\right)} 
\end{equation*}
\begin{equation}
    = \left(-s_2-i 0\right)^{\epsilon-1} \frac{1}{b} \frac{1}{\epsilon}\;\;{ }_2 F_1\left(1, \epsilon, 1+\epsilon ; 1- \frac{\left(s_2+a\right)}{b} \right) .
\end{equation}
In the second term, denoting $t = b y+a(1-y) $, one can organize the integral as 
\begin{equation}
\int_a^b d t=\int_0^b d t-\int_0^a d t=b \int_0^1 d x-a \int_0^1 d x \; .
\end{equation}
In this way, we obtain
\begin{equation*}
\int_0^1 \frac{d y}{s_2 (1-y) + b y+a (1-y)}(b y+a(1-y))^{\epsilon-1}=\frac{b^\epsilon}{s_2 b} \int_0^1 d x \frac{x^{\epsilon-1}}{1-x\left(1-\left(b-a\right) / s_2\right)} 
\end{equation*}
\begin{equation*}
-\frac{a^\epsilon}{s_2 b} \int_0^1 d x \frac{x^{\epsilon-1}}{1-x \left(1-(b-a)\left(s_2+a\right) /\left(s_2 b\right)\right)}=\frac{b^\epsilon}{s_2 b \epsilon}{ }_2 F_1\left(1, \epsilon, 1+\epsilon ; 1- \frac{(b-a)}{s_2} \right) 
\end{equation*}
\begin{equation}
-\frac{a^\epsilon}{s_2 b \epsilon}{ }_2 F_1\left(1, \epsilon, 1+\epsilon ; 1- \frac{(b-a)\left(s_2+a\right)}{\left(s_2 b\right)} \right) .
\end{equation}
Finally, restoring $t_1$ and $t_2$, we have
\begin{gather}
I_{4 B}=\frac{\pi^{2+\epsilon}}{s_2 t_2}  \frac{ \Gamma(1-\epsilon) \Gamma^2(1+\epsilon)}{\Gamma(1+2 \epsilon)} \frac{2}{\epsilon^2} \bigg[ \left(-s_2-i 0\right)^\epsilon { }_2 F_1\left(1, \epsilon, 1+\epsilon ; 1- \frac{\left(s_2-t_1 \right)}{(-t_2)} \right)\\  + (-t_2)^\epsilon { }_2 F_1\left(1, \epsilon, 1+\epsilon ; 1- \frac{(t_1-t_2)}{s_2} \right) - (-t_1)^{\epsilon} {}_2 F_1\left(1, \epsilon, 1+\epsilon ; 1- \frac{\left(s_2-t_1 \right)(t_1-t_2)}{s_2 (-t_2)} \right) \bigg] . \nonumber
\end{gather}
This result is exact and coincides with that in~(\ref{C1bis}).
\vspace{0.3 cm} \\
\textbf{Feynman integrals with logarithms} \vspace{0.3 cm} \\
In the calculation it happens to come across momentum integrals that have logarithms in the numerator. We explain below how to evaluate them, considering the example
\begin{equation}
    \int d^{D-2} k \frac{\ln \vec{k}^{\, 2} }{\vec{k}^{\, 2}  (\vec{k}+\vec{q}_2)^2} \;.
\end{equation}
Using $\ln \vec{k}^{\, 2}  = \frac{\partial}{\partial \alpha} (\vec{k}^{\, 2} )^{\alpha} \big |_{\alpha=0}$ and 
exchanging the order of integration and derivative, we get
\begin{equation*}
   \frac{\partial}{\partial \alpha} \left[ \int d^{D-2} k \frac{1}{(\vec{k}^{\, 2} )^{1-\alpha} (\vec{k}+\vec{q}_2)^2} \right]_{\alpha=0}  
\end{equation*}
\begin{equation*}
    = \pi^{1+\epsilon} (\vec{q}_2^{\; 2})^{-1+\epsilon} \frac{\partial}{\partial \alpha} \left[ (\vec{q}_2^{\; 2})^{\alpha} \frac{\Gamma(1-\alpha-\epsilon) \Gamma(\epsilon) \Gamma(\alpha+\epsilon)}{\Gamma(1-\alpha)\Gamma(2 \epsilon+\alpha)} \right]_{\alpha=0}
\end{equation*}
\begin{equation}
    = \pi^{1+\epsilon} (\vec{q}_2^{\; 2})^{-1+\epsilon} \frac{\Gamma(1-\epsilon) \Gamma^2 (\epsilon)}{\Gamma(2 \epsilon)} \bigg [ \ln \vec{q}_2^{\; 2} + \psi (1) + \psi (\epsilon) - \psi (1-\epsilon) - \psi (2 \epsilon) \bigg] \; .
\end{equation}

\section{Two-dimensional Euclidean and four-dimensional Minkowskian integrals}
\label{Appendix C}
In this Appendix, we give an explicit derivation of the following relations:
\begin{equation}
I_5 = \frac{\pi^{2+\epsilon} \Gamma (1-\epsilon)}{s} \left[ \ln \left( \frac{(-s) (\vec{q}_1-\vec{q}_2)^2}{(-s_1)(-s_2)} \right) \mathcal{I}_3 + \mathcal{L}_3 -\mathcal{I}_5 \right] \; ,
\tag{\ref{I5inTransver}}
\end{equation}
\begin{equation}
I_{4B}=- \frac{\pi^{2+\epsilon} \Gamma (1-\epsilon)}{s_2} \left[ \frac{\Gamma^2 (\epsilon)}{\Gamma (2 \epsilon)} (-t_2)^{\epsilon -1} \left( \ln \left( \frac{- s_2}{- t_2} \right) + \psi (1-\epsilon) - 2 \psi (\epsilon) + \psi (2 \epsilon)  \right) + \mathcal{I}_{4B}  \right] \; .
\tag{\ref{I4B-I4Bcal}}
\end{equation}

\vspace{0.2cm}
$\bullet$ Let us start from the definition of $I_5$:
\begin{equation*}
  I_{5} = \frac{1}{i} \int d^{D}k \frac{1}{(k^{2} + i \varepsilon) [(k+q_1)^{2}+i\varepsilon]
    [(k+q_2)^{2}+i\varepsilon] [(k+p_A)^2 + i \varepsilon]
    [(k-p_B)^2 + i \varepsilon]}
  \;.
\end{equation*}
We introduce the standard Sudakov decomposition for momenta,
\begin{equation}
    k = \alpha p_B + \beta p_A + k_{\perp} \; , \hspace{0.5 cm} d^{D} k = \frac{s}{2} d \alpha d \beta d^{D-2} k_{\perp} \; ,
\label{Sudakov1}
\end{equation}
\begin{equation}
    q_1 = p_A-p_{A'} = - \frac{\vec{q}_1^{\; 2}}{s} p_B + \frac{s_2}{s} p_A +q_{1 \perp} \; , \hspace{0.5 cm} q_2 = p_{B'}-p_{B} = - \frac{s_1}{s} p_B + \frac{\vec{q}_2^{\; 2}}{s} p_A + q_{2 \perp} \;.
\label{Sudakov2}
\end{equation}
By expressing denominators in terms of Sudakov variables, we can reduce the integration over $\alpha$ to a simple computation of residues in the complex plane. We have the following five simple poles:
\begin{equation}
    \alpha_1 = \frac{\vec{k}^{\, 2} - i \varepsilon}{\beta s} \; , \hspace{0.5 cm} \alpha_2 = \frac{(k+\vec{q}_1)^{2}-i \varepsilon}{s(\beta + \frac{s_2}{s})} + \frac{\vec{q}_{1}^{\; 2}}{s} \; , \hspace{0.5 cm} \alpha_3 = \frac{(k+\vec{q}_2)^{2}-i \varepsilon}{s(\beta + \frac{\vec{q}_{2}^{\; 2}}{s})} + \frac{s_1}{s} \; ,
\end{equation}
\begin{equation}
    \alpha_4 = \frac{\vec{k}^{2} - i \varepsilon}{(1+\beta) s} \; , \hspace{0.5 cm} \alpha_5 = \frac{\vec{k}^{2} - i \varepsilon}{\beta s} + 1 \; .
\end{equation}
We observe that in the region $\Omega = \{ \beta < -1 \; \;  \vee \; \; \beta > 0 \}$ the integral always vanishes, since all poles are on the same side with respect to the real $\alpha$-axis; 
in the region $\bar{\Omega} = \{ -1 < \beta < 0 \}$ we have contributions from three poles lying in the lower real $\alpha$-axis: $\alpha_4$ in the whole region $-1 < \beta < 0$, $\alpha_2$ in the region $-\frac{s_2}{s} < \beta < 0$ and $\alpha_3$ in the region $-\frac{\vec{q}_2^{\; 2}}{s} < \beta < 0$. The integral can therefore be expressed as
\begin{equation*}
\begin{split}
  I_5 = - \pi \int_{-1}^{0} \frac{d \beta}{1+\beta} & \int d^{D-2} k \frac{1}{\left[\alpha_4 \beta s - \vec{k}^{2} + i \varepsilon \right]} \frac{1}{\left[ (\alpha_4 - \frac{\vec{q}_{1}^{\; 2}}{s})(\beta + \frac{s_2}{s})s - (k+\vec{q}_1)^{2} + i \varepsilon \right]} \\ &
  \times \frac{1}{\left[(\alpha_4-1) \beta s - \vec{k}^{2} + i \varepsilon \right]} \frac{1}{\left[ (\alpha_4 - \frac{s_1}{s})(\beta + \frac{\vec{q}_2^{\; 2}}{s})s - (\vec{k}+\vec{q}_2)^{2} + i \varepsilon \right]}
 \end{split}
\end{equation*}
\begin{equation*}
\begin{split}
     \hspace{1.1 cm} - \pi \int_{-\frac{s_2}{s}}^{0} \frac{d \beta}{\beta + \frac{s_2}{s}} & \int d^{D-2} k \frac{1}{\left[\alpha_2 \beta s - \vec{k}^{2} + i \varepsilon \right]} \frac{1}{\left[ (\alpha_2 - \frac{s_1}{s})(\beta + \frac{\vec{q}_2^{\; 2}}{s})s - (\vec{k}+\vec{q}_2)^{2} + i \varepsilon \right]} \\ &
  \times \frac{1}{\left[\alpha_2 (1+\beta) s - \vec{k}^{2} + i \varepsilon \right]} \frac{1}{\left[(\alpha_2-1) \beta s - \vec{k}^{2} + i \varepsilon \right]} 
 \end{split}
\end{equation*}
\begin{equation}
\begin{split}
     \hspace{1.0 cm} - \pi \int_{-\frac{\vec{q}_2^{\; 2}}{s}}^{0} \frac{d \beta}{\beta + \frac{\vec{q}_2^{\; 2}}{s}} & \int d^{D-2} k \frac{1}{\left[\alpha_3 \beta s - \vec{k}^{2} + i \varepsilon \right]} \frac{1}{\left[ (\alpha_3 - \frac{\vec{q}_{1}^{\; 2}}{s})(\beta + \frac{s_2}{s})s - (\vec{k}+\vec{q}_1)^{2} + i \varepsilon \right]} \\ &
  \times \frac{1}{\left[\alpha_3 (1+\beta) s - \vec{k}^{2} + i \varepsilon \right]}  \frac{1}{\left[(\alpha_3-1) \beta s - \vec{k}^{2} + i \varepsilon \right]} \; .
 \end{split}
\end{equation}
Substituting the explicit values of poles and doing simple algebric manipulations, we end up with
(omitting the $i\varepsilon$'s in the denominators)
\begin{equation*}
    I_5 \simeq - \pi \int_0^1 d \beta \beta^3 \int d^{D-2} k \frac{1}{\vec{
    k}^{\, 2} (\vec{k} + \beta \vec{q}_1)^{2}[(\vec{k} + \beta \vec{q}_2)^{2} + \beta (1-\beta) (-s_1)][\vec{k}^{\, 2}  + \beta (1-\beta) (-s)]} 
\end{equation*}
\begin{equation*}
    \begin{split}
    + \pi \int_0^1 d \beta \beta^3 \int d^{D-2} k \left( \frac{s_2}{s} \right) & \frac{1}{(\vec{
    k}+\vec{q}_1)^{2} (\vec{k} +(1-\beta) \vec{q}_1 + \beta \vec{q}_2)^{2} [(\vec{k} + (1-\beta) \vec{q}_1)^{2}+\beta (1-\beta) \vec{q}_1^{\; 2}]} \\ 
    & \times \frac{1}{[(\vec{k}+(1-\beta) \vec{q}_1)^{2} + \beta (1-\beta) (-s_2)]}
    \end{split}
\end{equation*}
\begin{equation}
    - \pi \int_0^1 d \beta (1-\beta)^3 \int d^{D-2} k \frac{(\vec{q}_{2}^{\; 2})^2}{s_2 s} \frac{1}{[(\vec{k}+\vec{q}_2)^2]^2 (\vec{k} + \beta \vec{q}_2)^{2} [(\vec{k} + \beta \vec{q}_2)^{2}+\beta (1-\beta) \vec{q}_2^{\; 2}]} \; .
\end{equation}
The third integral is suppressed in the high-energy approximation; as for the other two, they can be calculated by first performing the change of variable $\vec k \to \beta \vec k$ in the integration over $\vec k$ and then integrating some terms in $\beta$ by using the following integral:
\begin{equation*}
    \int_0^1 d \beta \frac{(1-\beta)^{D-n}}{\delta + \beta} =  \int_0^1 d \beta \frac{(1-\beta)^{D-n}-1}{\delta + \beta} + \int_0^1 d \beta \frac{1}{\delta + \beta} 
\end{equation*}
\begin{equation}
    \simeq \int_0^1 d\beta \frac{(1-\beta)^{D-n}-1}{\beta} + \int_0^1 d \beta \frac{1}{\delta + \beta} \simeq \psi(1) - \psi(D-n-1) - \ln \delta \; ,
    \label{deltaTrick}
\end{equation}
where $\delta$ is a generic quantity tending to zero, like $\vec{k}^{\, 2} /s$ for instance, and $n=5$ and 6 in the cases we are interested in. We obtain
\begin{equation}
\begin{split}
    I_5 =& \; \frac{\pi}{s} \int d^{D-2} k \frac{1}{\vec{k}^{\, 2}  (\vec{k}+\vec{q}_1)^2(\vec{k}+\vec{q}_2)^2} \left[ \ln \left( \frac{(k+\vec{q}_2)^2 (-s)}{(-s_1)(-s_2)} \right) - ( \psi (1) -\psi (D-5)) \right] \\
    & + \pi \; \frac{\vec{q}_1^{\; 2}}{s} \int_0^1 d \beta \beta^{3} \int d^{D-2} k \frac{1}{\vec{k}^{\, 2}  (\vec{k}+\beta \vec{q}_1)^2 (\vec{k}+\beta \vec{q}_2)^2 [\vec{k}^{\, 2}  + \beta (1-\beta) \vec{q}_1^{\; 2}]} \; .
\end{split}
\end{equation}
To get the desired form, we rewrite the last two terms in the square bracket as an integral over $\beta$,
\[
\psi(1)-\psi(D-5) = \int_0^1 d\beta \, \frac{(1-\beta)^{D-6}-1}{\beta}
= \int_0^1 d\beta \, \frac{\beta^{D-6}}{1-\beta}-\int_0^1 d\beta \,\frac{1}{\beta} \;,
\]
and combine with the last term of the full expression, to get
\begin{equation*}
    I_5 = \; \frac{\pi}{s} \int d^{D-2} k \frac{1}{\vec{k}^{\, 2}  (\vec{k}+\vec{q}_1)^2(\vec{k}+\vec{q}_2)^2} \ln \left( \frac{(k+\vec{q}_2)^2 (-s)}{(-s_1)(-s_2)} \right) 
\end{equation*}
\begin{equation*}
    - \frac{\pi}{s} \hspace{-0.1 cm} \int_0^1 \hspace{-0.2 cm} \frac{d \beta}{1-\beta} \hspace{-0.1 cm} \int \hspace{-0.05 cm} \frac{d^{D-2} k}{\vec{k}^{\, 2} } \hspace{-0.05 cm} \left[ \frac{\beta^2}{(\vec{k}-\beta(\vec{q}_1-\vec{q}_2))^2[(1-\beta) \vec{k}^{\, 2}  + \beta (\vec{k}-\vec{q}_1)^2]} - \frac{1}{(\vec{k}-\vec{q}_1+\vec{q}_2)^2 (\vec{k}-\vec{q}_1)^2} \right] 
\end{equation*}
\begin{equation*}
     = \; \frac{\pi}{s} \int d^{D-2} k \frac{1}{\vec{k}^{\, 2}  (\vec{k}-\vec{q}_1)^2 (\vec{k}-\vec{q}_2)^2} \left[ \ln \left( \frac{(-s)(\vec{q}_1-\vec{q}_2)^2}{(-s_1)(-s_2)} \right) + \ln \left( \frac{(\vec{k}-\vec{q}_1)^2(\vec{k}-\vec{q}_2)^2}{\vec{k}^{\, 2}  (\vec{q}_1-\vec{q}_2)^{\; 2}} \right) \right]
\end{equation*}
\begin{equation*}
    - \frac{\pi}{s} \hspace{-0.1 cm} \int_0^1 \hspace{-0.2 cm} \frac{d \beta}{1-\beta} \hspace{-0.1 cm} \int \hspace{-0.05 cm} \frac{d^{D-2} k}{\vec{k}^{\, 2} } \hspace{-0.05 cm} \left[ \frac{\beta^2}{(\vec{k}-\beta(\vec{q}_1-\vec{q}_2))^2[(1-\beta) \vec{k}^{\, 2}  + \beta (\vec{k}-\vec{q}_1)^2]} - \frac{1}{(\vec{k}-\vec{q}_1+\vec{q}_2)^2 (\vec{k}-\vec{q}_1)^2} \right] 
\end{equation*}
\begin{equation*}
     + \frac{\pi}{s} \int d^{D-2} k \frac{1}{\vec{k}^{\, 2}  (\vec{k}-\vec{q}_1)^2 (\vec{k}-\vec{q}_2)^2} \left[ \ln \left( \frac{\vec{k}^{\, 2} }{(\vec{k}-\vec{q}_1)^2} \right) \right] \; .
\end{equation*}
Expressing the term in the last line as
    \begin{equation*}
     \frac{\pi}{s} \int_0^1 d \beta \frac{1}{1-\beta} \int d^{D-2} k \frac{1}{(\vec{k}-\vec{q}_1)^2 (\vec{k}-\vec{q}_2)^2} \left[ \frac{1}{\beta \vec{k}^{\, 2}  + (1-\beta) (\vec{k}-\vec{q}_1)^{\; 2}} - \frac{1}{\vec{k}^{\, 2} } \right] \, ,
\end{equation*}
and performing simple manipulations, we get
\begin{equation*}
\begin{split}
    I_5 &= \; \frac{\pi}{s} \int d^{D-2} k \frac{1}{\vec{k}^{\, 2}  (\vec{k}-\vec{q}_1)^2 (\vec{k}-\vec{q}_2)^2} \left[ \ln \left( \frac{(-s)(\vec{q}_1-\vec{q}_2)^2}{(-s_1)(-s_2)} \right) \right] \\ & + \frac{\pi}{s} \int d^{D-2} k \frac{1}{\vec{k}^{\, 2}  (\vec{k}-\vec{q}_1)^2 (\vec{k}-\vec{q}_2)^2} \left[\ln \left( \frac{(\vec{k}-\vec{q}_1)^2(\vec{k}-\vec{q}_2)^2}{\vec{k}^{\, 2}  (\vec{q}_1-\vec{q}_2)^{\; 2}} \right) \right]
\end{split}
\end{equation*}
\begin{equation*}
    - \frac{\pi}{s} \; \int_0^1 d \beta \frac{1}{1-\beta} \int d^{D-2} k \frac{1}{\vec{k}^{\, 2} [(1-\beta) \vec{k}^{\, 2}  + \beta (\vec{k}-\vec{q}_1)^2]}\left[ \frac{\beta^2}{(\vec{k}-\beta(\vec{q}_1-\vec{q}_2))^2} - \frac{1}{(\vec{k}-\vec{q}_1+\vec{q}_2)^2} \right] ,
\end{equation*}
which, by using definitions (\ref{I3cal}), (\ref{L3cal}), (\ref{I5cal}), leads exactly to eq.~(\ref{I5inTransver}). \\
\vspace{0.2cm}

$\bullet$ Let us prove the second relation; again we start from the definition of $I_{4B}$:
\begin{equation}
  I_{4B} = \frac{1}{i} \int d^{D}k \frac{1}{(k^{2} + i \varepsilon) [(k+q_1)^{2}+i\varepsilon]
    [(k+q_2)^{2}+i\varepsilon] [(k-p_B)^2 + i \varepsilon]}\;.
\tag{\ref{I4B}}
\end{equation}
We again introduce the Sudakov decomposition~(\ref{Sudakov1})-(\ref{Sudakov2}) and observe that we have four poles:
\begin{equation*}
   \beta_1 = \frac{\vec{k}^{\, 2} -i \varepsilon}{\alpha s} \; , \hspace{0.5 cm} \beta_2 = \frac{(\vec{k}+\vec{q}_1)^{2}-i \varepsilon}{(\alpha-\frac{\vec{q}_1^{\; 2}}{s}) s} - \frac{s_2}{s} \; , \hspace{0.5 cm} \beta_3 = \frac{(\vec{k}+\vec{q}_2)^{2}-i \varepsilon}{(\alpha-\frac{s_1}{s})s} - \frac{\vec{q}_2^{\; 2}}{s} \; , \hspace{0.5 cm}  \beta_4 = \frac{\vec{k}^{\, 2} -i \varepsilon}{(\alpha-1)s} \; .
\end{equation*}
In the region $\Omega = \{ \alpha < 0 \; \;  \vee \; \; \alpha > 1 \}$ the integral always vanishes, while, in the region $\bar{\Omega} = \{  0 < \alpha < 1\}$, if we close the integration path over $\beta$ in the half-plane $\Im \beta > 0$ and use the residue theorem, we have
\begin{equation*}
I_{4 B} = -\pi \int_0^1 \frac{d \alpha}{1-\alpha} \int d^{D-2} k \ \frac{1}{(\alpha \beta_4 s - \vec{k}^{2} + i \varepsilon )}
\end{equation*}
\begin{equation*}
\times \frac{1}{[\alpha (\beta_4 + \frac{s_2}{s})s - (\vec{k}+\vec{q}_1)^2 + i \varepsilon][ (\alpha-\frac{s_1}{s}) (\beta_4 + \frac{\vec{q}_2^{\; 2}}{s})s - (\vec{k}+\vec{q}_2)^2 + i \varepsilon]}
\end{equation*}
\begin{equation}
    + \pi \int_0^{\frac{s_1}{s}} \frac{d \alpha}{\alpha-\frac{s_1}{s}} \int \frac{d^{D-2} k}{(\alpha \beta_3 s - \vec{k}^{2} + i \varepsilon )[\alpha (\beta_3 + \frac{s_2}{s})s - (\vec{k}+\vec{q}_1)^2 + i \epsilon][ (\alpha-1) \beta_3 s - \vec{k}^2 + i \varepsilon]} \; .
\end{equation}
In the last expression, we neglected the contribution of the pole $\beta_2$, which is present in the region $ \Omega' = \{ 0<\alpha <\vec{q}_1^{\; 2}/s \}$, since it is suppressed in the MRK. For the same reason, in the previous integrals we approximated $\alpha- \vec q_1^{\:2}/s$ with $\alpha$.
Substituting the explicit values of $\beta_3$ and $\beta_4$, one gets (recalling that 
$s_1 s_2/s = (\vec q_1-\vec q_2)^2$ and taking the high-energy limit)
\begin{equation}
    I_{4B} = \pi \int_0^1 d \alpha  \alpha^2 \int d^{D-2} k \frac{1}{\vec{k}^{2} (\vec{k}+\alpha \vec{q}_2)^2 [(\vec{k}+\alpha \vec{q}_1)^2+\alpha(1-\alpha)(-s_2-i \varepsilon)]}
\end{equation}
\begin{equation*}
    - \pi \frac{(\vec{q}_1-\vec{q}_2)^2}{s_2} \int_0^1 \hspace{-0.1 cm} d \alpha \; \alpha^2 \int \frac{ d^{D-2} k}{(\vec{k}+\vec{q}_2)^2[(\vec{k}+(1-\alpha) \vec{q}_2)^2 + \alpha (1-\alpha) \vec{q}_2^{\; 2}](\vec{k}+(1-\alpha) \vec{q}_2 + \alpha \vec{q}_1)^2} \; .
\end{equation*}
Manipulating the first term and using the integral in eq.~(\ref{deltaTrick}), one can obtain the following form:
\begin{equation*}
I_{4 B} = \frac{\pi^{2+\epsilon}}{s_2 t_2} \frac{\Gamma (1-\epsilon) \Gamma^2 (\epsilon)}{\Gamma (2 \epsilon)} (\vec{q}_2^{\; 2})^{\epsilon} \left[ \ln \left( \frac{-s_2}{-t_2} \right) + \psi (1-\epsilon) - \psi (\epsilon) \right] 
\end{equation*}
\begin{equation*}
    + \frac{\pi}{s_2} \int_0^1 \frac{d \alpha}{\alpha} \int d^{D-2} k \frac{1}{(\vec{k}-\vec{q}_1)^2 (\vec{k}-(\vec{q}_1-\vec{q}_2))^2}   
\end{equation*}
\begin{equation*}
    -\frac{\pi}{s_2} \int_0^1 \frac{d \alpha}{1-\alpha} \int d^{D-2} k \frac{1}{(\vec{k}+\vec{q}_2)^2 \left[ \alpha \vec{k}^2 + (1-\alpha) (\vec{k}+\vec{q}_1)^{2} \right]} 
\end{equation*}
\begin{equation}
     - \pi \frac{(\vec{q}_1-\vec{q}_2)^2}{s_2} \int_0^1 d \alpha \; \alpha^2 \int d^{D-2} k \frac{1}{\vec{k}^2[(\vec{k}+\alpha \vec{q}_2)^2 + \alpha (1-\alpha) \vec{q}_2^{\; 2}](\vec{k} + \alpha (\vec{q}_2 - \vec{q}_1))^2} \; .
\end{equation}
In the last integral we perform the transformation 
\begin{equation*}
    \vec{k} \rightarrow \vec{k}' = \vec{k} - \vec{p} \; , \hspace{0.5 cm} \alpha \rightarrow \alpha ' = \frac{\alpha \vec{k'}^{2}}{(1-\alpha) \vec{k}^{2} + \alpha \vec{k'}^{2}} \; \; \; \; {\rm{with}} \; \; \; \; \vec{p} \equiv \vec{q}_1-\vec{q}_2 \;,
\end{equation*}
and get
\begin{equation*}
I_{4B} = \frac{\pi^{2+\epsilon}}{s_2 t_2} \frac{\Gamma (1-\epsilon) \Gamma^2 (\epsilon)}{\Gamma (2 \epsilon)} (\vec{q}_2^{\; 2})^{\epsilon} \left[ \ln \left( \frac{-s_2}{-t_2} \right) + \psi (1-\epsilon) - \psi (\epsilon) \right]  
\end{equation*}
\begin{equation*}
    + \frac{\pi}{s_2} \int_0^1 \frac{d \alpha}{\alpha} \int d^{D-2} k \frac{1}{(\vec{k}-\vec{q}_1)^2 (\vec{k}-(\vec{q}_1-\vec{q}_2))^2}  
\end{equation*}
\begin{equation*}
    -\frac{\pi}{s_2} \int_0^1 \frac{d \alpha}{1-\alpha} \int d^{D-2} k \frac{1}{(\vec{k}+\vec{q}_2)^2 \left[ \alpha \vec{k}^2 + (1-\alpha) (\vec{k}+\vec{q}_1)^{2} \right]} 
\end{equation*}
\begin{equation}
     - \frac{\pi}{s_2} \int_0^1 d \alpha \; \int d^{D-2} k \frac{ \alpha^2 \vec p^{\; 2}} {(\vec{k} + \alpha \vec{p})^2 [(\vec{k}+\alpha \vec{p})^2 + \alpha (1-\alpha) \vec{p}^{\; 2}] [ (1-\alpha) \vec{k}^2 + \alpha (\vec{k} + \vec{q}_1)^2 ]} \; .
\end{equation}
The last two terms can be shown to lead to 
\begin{equation*}
    - \frac{\pi}{s_2} \int_{0}^1 \frac{d \alpha}{\alpha} \int d^{D-2} k \frac{1-\alpha}{(\vec{k}-(1-\alpha) (\vec{q}_1-\vec{q}_2))^2 [\alpha \vec{k}^{\, 2}  + (1-\alpha) (\vec{k}-\vec{q}_1)^2 ]} 
\end{equation*}
\begin{equation}
    + \frac{\pi^{2+\epsilon}}{s_2 t_2} \frac{\Gamma (1-\epsilon) \Gamma^2 (\epsilon)}{\Gamma (2 \epsilon)} (\vec{q}_2^{\; 2})^{\epsilon} (\psi (2 \epsilon) - \psi (\epsilon)) \; ,
\end{equation}
so that we finally find
\begin{equation*}
I_{4B}=- \frac{\pi^{2+\epsilon} \Gamma (1-\epsilon)}{s_2} \left[ \frac{\Gamma^2 (\epsilon)}{\Gamma (2 \epsilon)} (-t_2)^{\epsilon -1} \left( \ln \left( \frac{- s_2}{- t_2} \right) + \psi (1-\epsilon) - 2 \psi (\epsilon) + \psi (2 \epsilon)  \right) \right] 
\end{equation*}
\begin{equation*}
 - \frac{\pi^{2+\epsilon} \Gamma (1-\epsilon)}{s_2} \int_0^1 \frac{d \alpha}{ \alpha} \int \frac{d^{D-2} k}{\pi^{1 + \epsilon} \Gamma (1-\epsilon)}  
\end{equation*}
\begin{equation}
   \times \left[ \frac{1-\alpha}{\left[ \alpha \vec{k}^2 + (1- \alpha) (\vec{k}-\vec{q}_1)^2 \right] (\vec{k}-(1-\alpha)(\vec{q}_1-\vec{q}_2))^2} - \frac{1}{(\vec{k}-\vec{q}_{1})^2 (\vec{k}-(\vec{q}_1-\vec{q}_2))^2} \right] \; ,
\end{equation}
which is exactly eq.~(\ref{I4B-I4Bcal}).

\section{Soft limit}
\label{Appendix D}
In this Appendix we evaluate the soft limit ($\vec{p} \to 0$) of the integrals considered so far. 
We start from $\mathcal{I}_3$, given by eq.~(\ref{I3cal}). In the soft limit, the dominant contribution comes from the region $ \vec{k} \simeq \vec{q}_1 \simeq \vec{q}_2$; hence we can make the replacement
\begin{equation*}
    \frac{1}{\vec{k}^{\, 2} } \to  \frac{1}{\vec{Q}^{\; 2}} \hspace{0.5 cm} \text{with} \hspace{0.5 cm} \vec{Q} \equiv \frac{\vec{q}_1+\vec{q}_2}{2}
\end{equation*}
and obtain 
\begin{equation}
    \mathcal{I}_3 \simeq \frac{1}{\vec{Q}^2} \frac{1}{\pi^{1+\epsilon} \Gamma (1 - \epsilon)} \int d^{2+2 \epsilon} k \frac{1}{\vec{k}^{\, 2}  (\vec{k}+\vec{p})^2} = \frac{1}{\vec{Q}^2} \frac{\Gamma^2 (\epsilon)}{\Gamma (2 \epsilon)} (\vec{p}^{\; 2})^{\epsilon-1} \;.
    \label{I3calSoft}
\end{equation}
To obtain the soft limit of $\mathcal{I}_{4B}$, eq.~(\ref{I4Bcal}), we start from the representation
\begin{equation*}
    \mathcal{I}_{4B} = \int_0^1 \frac{dx}{x} (J_B(x)-J_B(0)) \; ,
\end{equation*}
where
\begin{equation}
    J_B (x) = \int_0^1 dz \frac{1-x}{\left[ z(1-x)(a x + b (1-x) (1-z)) \right]^{1-\epsilon}} 
    \;, \;\;\;\;\; a= \vec q_1^{\;2}, \; b= \vec q_2^{\;2} \; ,
    \label{JB}
\end{equation}
and we observe that, in the soft region
\begin{equation*}
    \vec{q}_1 \simeq \vec{q}_2 \simeq \frac{\vec{q}_1 + \vec{q}_2}{2} = \vec{Q} \; ,
\end{equation*}
$J_B(x)$ becomes
\begin{equation*}
    J_B(x) \simeq \frac{1}{(\vec{Q}^{2})^{1-\epsilon}} \int_0^1 dz \frac{1}{z^{1-\epsilon} (1-x)^{-\epsilon} [1-z(1-x)]^{1-\epsilon}} = \frac{1}{(\vec{Q}^{2})^{1-\epsilon}} \int_0^{1-x} dz \ [z (1-z)]^{\epsilon-1} 
\end{equation*}
and 
\begin{equation*}
    \mathcal{I}_{4B} = - \frac{1}{(\vec{Q}^{2})^{1-\epsilon}} \int_0^1 \frac{dx}{x} \int_{1-x}^1 dz \; [z  (1-z)]^{\epsilon-1} = - \frac{1}{(\vec{Q}^{2})^{1-\epsilon}} \int_0^1 dz \ [z  (1-z)]^{\epsilon-1} \int_{1-z}^1 \frac{dx}{x}
    \end{equation*}
    \begin{equation}
    = \frac{\Gamma^2(\epsilon)}{\Gamma (2 \epsilon)} \frac{1}{(\vec{Q}^{2})^{1-\epsilon}} (\psi (\epsilon) - \psi (2 \epsilon)) \;.
    \label{I4BcalSoft}
\end{equation}
The soft limit of the pentagon integral, eq.~(\ref{I5}), can be immediately obtained and reads~\cite{Fadin:2000yp}
\begin{equation}
    I_5 \simeq \frac{\pi^{2+\epsilon} \Gamma (1-\epsilon)}{s} \frac{\Gamma^2 (\epsilon)}{\Gamma (2 \epsilon)} \frac{(\vec{p}^{\; 2})^{\epsilon-1}}{\vec{Q}^{2}} (\psi (\epsilon) - \psi (1-\epsilon) + i \pi) \; .
\end{equation}
From this, by using eq.~(\ref{I5inTransver}), we obtain
\begin{equation}
    \mathcal{I}_5-\mathcal{L}_3 \simeq  \frac{\Gamma^2 (\epsilon)}{\Gamma (2 \epsilon)} \frac{(\vec{p}^{\; 2})^{\epsilon-1}}{\vec{Q}^{2}} ( \psi (1-\epsilon) - \psi (\epsilon)) \;.
    \label{I5cal-L3calSoft}
\end{equation}

\clearpage

\section*{Acknowledgments}

We thanks Marco Rossi, Lech Szymanowski and Samuel Wallon for useful discussions. \\ M.F. thanks IJCLab for support while part of this work was done.  \\ 
M.F. and A.P acknowledge support from the INFN/QFT@COL\-LI\-DERS project.

\bibliographystyle{apsrev}
\bibliography{references}

\end{document}